\documentclass{aa}  
\usepackage{graphicx}
\usepackage{color}
\usepackage{txfonts}
\begin{document}

        \title{Galaxy evolution in merging clusters: The passive core of the "Train Wreck" cluster of galaxies, \object{A520}\thanks{The full table with measured redshifts is only available in electronic form at the CDS via anonymous ftp to cdsarc.u-strasbg.fr (130.79.128.5) or via http://cdsweb.u-strasbg.fr/cgi-bin/qcat?J/A+A/}}
   \subtitle{}
   \author{Boris Deshev \inst{\ref{inst1},\ref{inst2}}
                \and Alexis Finoguenov \inst{\ref{inst3},\ref{inst4}}
                \and Miguel Verdugo \inst{\ref{inst5}}
                \and Bodo Ziegler \inst{\ref{inst5}}
                \and Changbom Park \inst{\ref{inst6}}
                \and Ho Seong Hwang \inst{\ref{inst7}}
                \and Christopher Haines \inst{\ref{inst8}}
                \and Peter Kamphuis \inst{\ref{inst9}}
                \and Antti Tamm \inst{\ref{inst1}}
                \and Maret Einasto \inst{\ref{inst1}}
                \and Narae Hwang \inst{\ref{inst10}} 
                \and Byeong-Gon Park \inst{\ref{inst10}}
          }
   \institute{Tartu Observatory, T\~oravere, 61602, Estonia \\
                        \email{boris.deshev@to.ee}\label{inst1}
                \and
                        Institute of Physics, University of Tartu, Ravila 14c, 50411, Estonia\label{inst2}
                \and
                        Department of Physics, University of Helsinki, PO Box 64, FI-00014 Helsinki, Finland\label{inst3}
                \and
                        Max-Planck-Institute for Extraterrestrial Physics, Giessenbachstraße, 85748 Garching, Germany\label{inst4}             
                \and                    
                        Institute for Astronomy (IfA), University of Vienna,
              T\"urkenschanzstrasse 17, A-1180 Vienna\label{inst5}
                \and
                        School of Physics, Korea Institute for 
Advanced Study, 85 Hoegiro, Dongdaemun-gu, Seoul 02455, Republic of Korea\label{inst6}                  
                \and
                Quantum Universe Center, Korea Institute for Advanced 
Study, 85 Hoegiro, Dongdaemun-gu, Seoul 02455, Republic of Korea\label{inst7}
        \and
                INAF - Osservatorio Astronomico di Brera, via Brera 28, 20122 Milano,
                        via. E. Bianchi 46, 23807 Merate, Italy\label{inst8}
        \and
                National Centre for Radio Astrophysics, TIFR, Ganeshkhind, Pune 411007, India\label{inst9}
                \and
                        Korea Astronomy and 
Space Science Institute, 776 Daedeokdae-Ro, Yuseong-Gu, Daejeon 34055, 
Republic of Korea \label{inst10}            
             }
   \date{Received ; accepted }
 
  \abstract
   {}
   {The mergers of galaxy clusters are the most energetic events in the universe after the Big Bang. With the increased availability of multi-object spectroscopy and X-ray data, an ever increasing fraction of local clusters are recognised as exhibiting signs of recent or past merging events on various scales. Our goal is to probe how these mergers affect the evolution and content of their member galaxies. We specifically aim to answer the following questions: Is the quenching of star formation in merging clusters enhanced when compared with relaxed clusters? Is the quenching preceded by a (short-lived) burst of star formation?}
   {We obtained optical spectroscopy of $>400$ galaxies in the field of the merging cluster Abell 520. We combine these observations with archival data to obtain a comprehensive picture of the state of star formation in the members of this merging cluster.  Finally, we compare these observations with  a control sample of ten non-merging clusters at the same redshift from The Arizona Cluster Redshift Survey (ACReS). We split the member galaxies into passive, star forming or recently quenched depending on their spectra.}
   {The core of the merger shows a decreased fraction of star forming galaxies compared to clusters in the non-merging sample. This region, dominated by passive galaxies, is extended along the axis of the merger. We find evidence of rapid quenching of the galaxies during the core passage with no signs of a star burst on the time scales of the merger ($\lesssim$0.4~Gyr).
   Additionally, we report the tentative discovery of an infalling group along the main filament feeding the merger, currently at ${\sim}2.5$ Mpc from the merger centre. This group contains a high fraction of star forming galaxies as well as approximately two thirds of all the recently quenched galaxies in our survey.}
   {}
   \maketitle
%
\section{Introduction}
Galaxies are not uniformly distributed throughout the universe. The large-scale structure of the universe \citep{JoeveerEinasto1978} is a mixture of different environments in which galaxies are formed and are evolving. The debate as to when the effects of environments were more decisive for the currently observed properties of the galaxies is termed nature versus nurture \citep[see e.g.][]{Bamford2009,Peng2010}. It has long been known that  the mixture of galaxy types varies significantly with increasing density of the environment \citep{Hubble&Humason1931}. The low-density environments are populated primarily by gas-rich, star forming galaxies with late-type morphologies, while the densest parts of the clusters are almost completely devoid of any cold gas or signs of star-formation, and are dominated by early-type galaxies \citep[e.g.][]{Dressler1980, Dressler1997, Solanes2001, Lewis2002, Gomez2003, Jaffe2016}. 

        Studies of the environmental effects on galaxy evolution measure environment in a variety of ways which are thus sensitive to different physics. Some mechanisms, like ram-pressure stripping \citep{Gunn&Gott1972} from the high-velocity interaction between the galaxies' inter-stellar medium (ISM) and the hot intra-cluster matter (ICM), are expected to only affect galaxies when they become satellites in relatively massive haloes where they can strip their gaseous halo and the outskirts of the cold gas disks. An increase in local galaxy density is most likely to affect galaxies via tidal interactions or mergers \citep{Spitzer&Baade1951, Boselli&Gavazzi2006}. The properties of the nearest neighbour also have an effect \citep{ParkHwang2009,Kauffmann2013}, in the sense that neighbouring galaxies usually have similar star formation properties.

        The hierarchical structure formation in a $\Lambda$CDM universe infers a hierarchy of environments for the galaxies. Traditionally this ranges from the lowest-density voids, through the small groups and outskirts of large clusters, to the interiors of the largest virialised clusters.  Beyond that, the hierarchy of environments continues with conglomerates of clusters - "superclusters" and merging clusters. The effects of this high-end environment are still poorly understood. There are signs that the environment in which the massive clusters reside (superclusters) also has an effect on their constituent galaxies, as the superclusters with complicated inner structure tend to host clusters with a  higher fraction of star forming galaxies than superclusters with simple inner structure \citep{Einasto2014}.

The fraction of clusters experiencing major mergers is likely significantly larger than commonly perceived \citep{McGee2009, Cassano2016}. Observational confirmation of the unsettled nature of a cluster requires specific conditions and is likely possible for a limited amount of time. The angle between the merger axis and the plane of the sky should be small enough so that an angular separation is visible. This, however, will make the velocity segregation smaller. X-ray observations have to "catch" the merger before the shock moves to the outer, low-surface-brightness regions \citep{Markevitch2005}, but also a small inclination angle would make it easier to detect the density increase. Because of this, cluster mergers with a mass ratio of less than a few are exceedingly rare and detailed analyses of their galaxy populations have only been performed in recent times \citep[e.g.][]{Abraham1996, Poggianti2006, ChungSM2009, HwangLee2009, Shim2011, Sobral2015}. Their effects are expected to be akin to the effects that virialised clusters have on their constituent galaxies, although both ram pressure stripping and tidal interaction can be significantly strengthened in merging clusters due to the usually very high velocities involved, the structured ICM and the clumpy gravitational potential \citep{TonnesenBryan&vanGorkom2007, Tonnesen&Bryan2008, Kapferer2008}. There is still an ongoing debate as to whether the onset of environmental effects, which ultimately quenches star formation, is accompanied by initial, short lived star burst \citep{Vollmer2001, Bekki&Couch2003, Fujita&Nagashima1999, Poggianti2008, HendersonBekki2016}. One additional physical mechanism acting on galaxies in merging clusters is the shock waves propagating through the ICM \citep{Roediger2014}. If a galaxy still contains a sufficient amount of cold gas, it can be compressed by the shock and trigger a star burst lasting a few hundred million years, until the depletion of the available gas \citep{Owers2012, Ebeling2014, Pranger2014, Stroe2015}. Recently \citet{Stroe2015} observed this remarkable rejuvenation of some of the members of the CIZA J2242.8+5301 (named the "Sausage") cluster, a relatively young merger  <0.7 Gyr old.

        In this work we embark on a survey of the properties of the galaxies in another well-known cluster merger \citep{Struble1987,Markevitch2005,Mann&Ebeling2012,Wang2016}- Abell 520 (A520 hereafter) also known as the "Train Wreck" cluster. 
        
        This paper is organised as follows: Section~\ref{a520} describes our target cluster and summarises the results from previous studies of this system, Section~\ref{data} describes the data acquisition, processing and analysis, Section~\ref{gal_class} defines the galaxy classification scheme we are using in the following sections, and Section~\ref{results} presents the first scientific results from our survey. In Section~\ref{discussion} we  discuss the interpretation of our results and compare them with other studies of merging clusters of galaxies. In Section 7 we end by summarising our findings.
        
        Throughout, we use a standard $\Lambda$CDM cosmology with H$_\mathrm{0}=70$ km s$^{-1}$ Mpc$^{-1}$, $\Omega_\mathrm{m}=0.3$, and $\Omega_\mathrm{\Lambda}=0.7$, thus 1\arcsec corresponds to 3.3 kpc at the redshift of A520.

\section{The merging cluster Abell 520}\label{a520}
        A520 is a massive merger of at least two sub clusters at z=0.201. The latest mass estimate from weak lensing \citep{Hoekstra2015} estimates its virial mass M$_\mathrm{vir}=15.3\pm3 \times10^{14}$M$_{\odot}$. The mass ratio of the main merging components is 1.0 \citep{Jee2014}. So far this cluster has received a lot of attention due to its peculiar mass distribution. There have been no fewer than six separate weak lensing mass reconstructions based on four separate data sets (\citet{Mahdavi2007}/CFHT, \citet{Okabe&Umetsu2008}/SUBARU, \citet{Jee2012}/HST+WFC, \citet{Clowe2012}/HST+ACS, \citet{Jee2014}/HST+ACS, \citet{Hoekstra2015}/CFHT). Four of the above detect a mass concentration near the centre of the merger, coinciding with the peak of X-ray emission, with unusually high mass to light ratio M/L${\simeq}$800 (\citet{Jee2014}; approximately four times the M/L ratio of the rest of the cluster). \citet{Clowe2012} does not detect this mass concentration and \citet{Hoekstra2015} do not provide mass maps and do not comment on its presence. All these studies show a picture of two sub-clusters observed after the core passage and currently separated by  ${\sim}4.5 \arcmin$ on the sky (${\sim}0.8$Mpc) along a NE-SW axis. These sub-clusters are named P2 and P4 by \citet{Jee2012} and we use these names throughout this article. Their positions are indicated on Fig.~\ref{SFR_on_the_sky}. The hot intra-cluster gas is stripped from these sub-clusters and resides between them. The X-ray emission shows a shock front which gives a Mach number of the merger of $2.1^{+0.4}_{-0.3}$\citep{Markevitch2005}. Using this, \citet{Mahdavi2007} calculate the time since the core passage as ${\sim}1$ Gyr and an inclination with regards to the plane of the sky of ${\sim}60^{\circ}$, although from dynamical considerations \citet[][G08 hereafter]{Girardi2008} estimate a much shorter time scale and lower inclination of 0.2-0.3~Gyr and ${\sim}30^{\circ}$, respectively. The weak-lensing mass maps also show several other mass concentrations lying to the east and west of the merger, and one to the northeast of P2 (see e.g. Fig. 5 in \cite{Jee2014} and Fig. 1 in G08). The detailed analysis of deep Chandra X-ray imaging of A520 by \citet{Wang2016} also paints a picture of a very complex merger.  Because of this complexity, A520 has been extensively studied in terms of alternative gravity and dark matter theories \citep[e.g.][]{Moffat2009, Specthmann2016}. G08 presented an exhaustive analysis of the velocity distribution of 293 galaxies within an irregularly shaped region of approximately 10\arcmin$\times$10\arcmin at the centre of the merger, 167 of which are cluster members. They confirm the complex nature of the merger and speculate that A520 could potentially be residing at the crossing of three large-scale structure filaments, two lying approximately in the plane of the sky and a third one aligned with the line of sight thus giving rise to the apparent dark matter concentration discussed above. All this attention to A520 is concentrated in a very small region on the sky, only partially extending out to the virial radius of the cluster.
        
        While A520 as a merger, or a cluster in formation, has received a lot of attention, its galaxy population has so far escaped such scrutiny. A520 is part of the \citet{Butcher&Oemler1984} sample. Notably it has the lowest blue fraction from their sample at $z{\sim}0.2$. It is interesting to compare it with A963 (part of our comparison sample), the cluster with the highest blue fraction at that redshift ($f_{B}$=19$\pm$5\%, \cite{Butcher&Oemler1984}). Recent research \citep{Jaffe2016} has shown that A963 contains a large amount of substructure. Perhaps the reason for the significantly different blue fractions in these two clusters is the recent violent merger of equal mass subclusters that A520 has experienced, as opposed to the continuous accretion of smaller groups that build up A963 over time.
        
        What makes A520 interesting for us is the fact that it is one of the lowest redshift mergers of subclusters of equal mass \citep{Mann&Ebeling2012} that show a clear separation between galaxies and ICM. In addition, the time scale of the merger is reasonably well constrained from X-ray observations \citep{Markevitch2005} allowing us to test for the presence of merger-induced star formation.

\section{Data}\label{data}
\subsection{Archival data}\label{additional_data}
\subsubsection{Imaging}\label{add_imaging}
        In this analysis we made use of $g'$ and $r'$ wide field imaging of 1 square degree around A520 using Megacam at the Canada France Hawaii Telescope (CFHT). The data were collected in November 2004 (proposal ID:04BC99, PI Henk Hoekstra) and processed with MegaPipe \citep{Gwyn2008}. We downloaded the reduced and photometrically calibrated mosaics
from the archive. We performed photometry with SExtractor \citep{Bertin1996} using MAG\_AUTO magnitudes. SExtractor was run in dual mode to ensure the measurements are performed over the same apertures. The star/galaxy separation was based on a total magnitude - half light radius plot \citep{Erben2005}, complemented with a visual inspection by BD. The $g'$ and $r'$ magnitudes were corrected for Galactic extinction according to \citet{Schlafly2011}. A K-correction calculated with the ``K-corrections calculator'' by \citet{Chilingarian2010} was applied.
\subsubsection{Spectroscopy}\label{add_spect}
        G08 published the redshifts of 293 galaxies in the field of A520, combining new observations with archival data presented by \cite{Yee1996} and \cite{Proust2000}. For part of our analysis (presented in Sections~\ref{glob_params} and \ref{substructure}) we combine these data with the MMT data presented in Section~\ref{MMTspec}. We refer to this combined data set as \textit{"full data set"}. The rest of the analysis presented here is based on the \textit{"MMT data set"} (Section~\ref{MMTspec}) alone, for it is the only one providing information on the presence/absence of emission lines, allowing us to do spectral classification of the galaxies.
\subsubsection{X-ray data}\label{add_xray}
        In this work we use XMM-Newton X-ray imaging of A520. The observations were done in Sept, 2004 (ObsID: 0201510101, PI: T. Clarke). 
        
\subsection{New spectroscopic data}\label{MMTspec}
        This paper presents new observations from the multi-fibre spectrograph "Hectospec" \citep{Fabricant2005} mounted on the 6.5m Multi-Mirror Telescope (MMT\footnote{Observations reported here were obtained at the MMT Observatory, a joint facility of the University of Arizona and the Smithsonian Institution.}) in Arizona, USA to measure accurate redshifts of galaxies in and around A520 and to estimate the properties of their ISM and stellar populations.
\subsubsection{Target selection}\label{target_selection}
\begin{figure}
   \centering
   \includegraphics[width=\hsize]{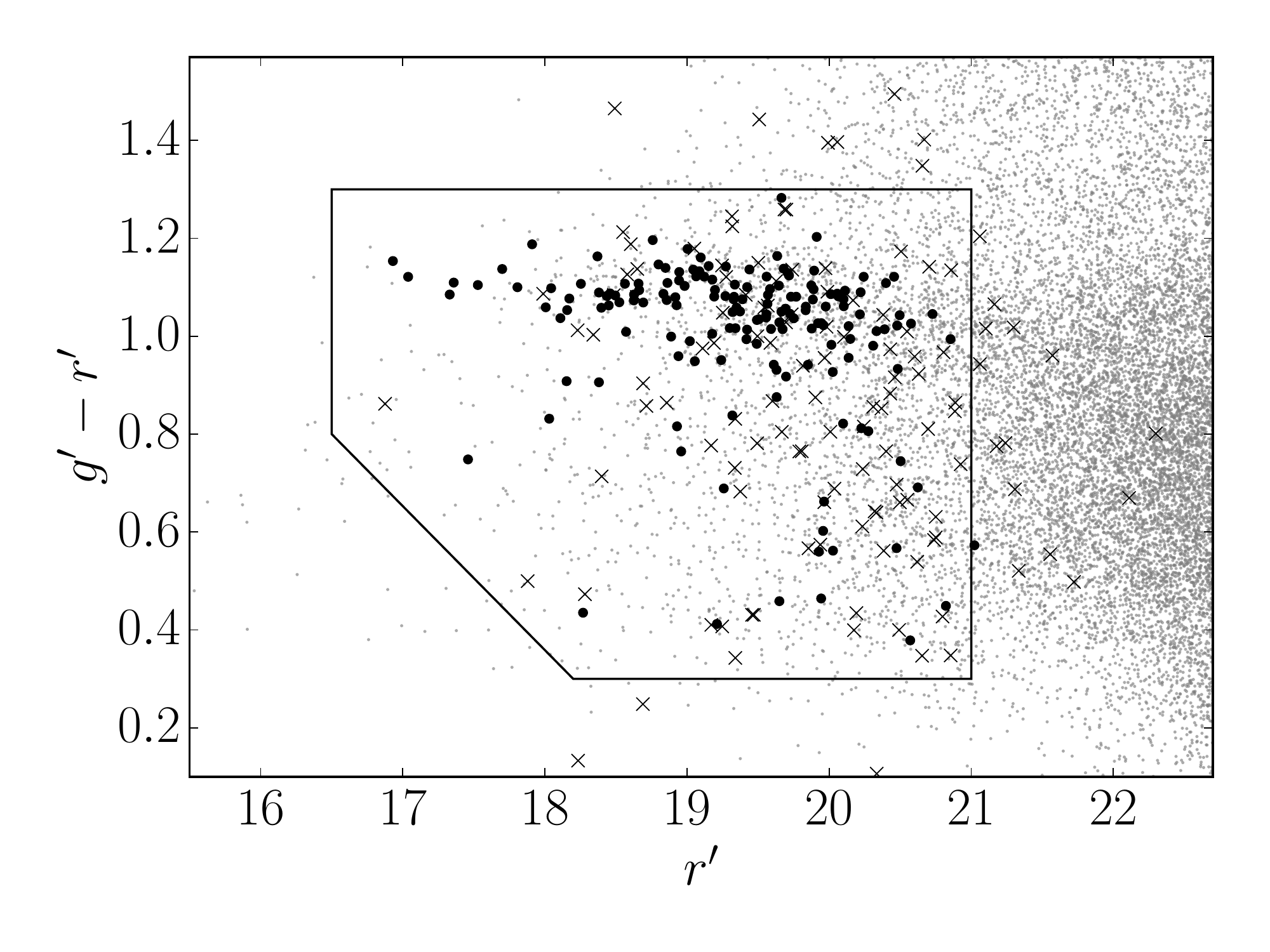}
      \caption{Selection of targets by optical colour and magnitude. The grey points are all the extended sources extracted from the CFHT images. The black dots and crosses are the spectroscopically confirmed members and non-members, respectively. The black polygon shows the region from which we select our targets.}
         \label{col_mag}
   \end{figure}

        The target selection was based on the $g'$ and $r'$ band images (see Section~\ref{add_imaging}). Prior to target selection the two images were astrometrically calibrated using the positions of point sources from the 2MASS \citep{Skrutskie2006} survey. Guide stars for the observations were selected from the same survey.
        The first step is based on the colour-magnitude diagram shown in Fig.~\ref{col_mag}. The grey points show all the extended sources extracted from the CFHT images. The dots and crosses are the members and non-members identified as such by G08. The black polygon shows the region from which we draw our sources. The faint magnitude limit is set equal to the completeness limit of the currently available spectral data. This is also the magnitude limit of the \citet{Rines2013} survey performed with the same telescope and instrument combination, which we use as a guidance for the instrument performance. One hour integration provides sufficient signal to noise ratio in the continuum for redshift estimation of targets with $r' < 21$. The colour selection is $0.3 < g'-r' < 1.3$, which encompasses the entire red sequence and the blue cloud of the confirmed cluster members. In order to remove any foreground contamination we limit the selected sources by their angular size on the sky, selecting sources with 50\% flux radius (R$_\mathrm{50}$) smaller than $1.25 \times R_\mathrm{50}$ of the largest confirmed cluster member. All the targets have sizes on the sky significantly exceeding the fibre diameter. We remove all the confirmed non-members from the target list.
        After applying these criteria we end up with 2523 potential targets, to which we assign priorities which are aimed at maximising the fraction of cluster members from the observed galaxies. We decrease the priority of potential targets residing in a region with low fraction of cluster members among the previously published data. Additionally we decrease the priority for galaxies by one if their magnitude within a central aperture with a diameter of 1.5\arcsec is below 20\% of the dark sky brightness, and by 2 if it is below 20\% of the grey sky brightness \citep[see ][for description of the instrument capabilities]{Fabricant2005}. The priority is also decreased for targets already observed by G08. Targets at a cluster-centric distance $r>3$Mpc were also given lower priorities. 
        The so-produced list of targets is fed into the fibre assignment program \textit{xfitfibs} (see Hectospec observing manual\footnote{https://www.cfa.harvard.edu/mmti/hectospec/hecto\_software\_manual.htm}).
\subsubsection{Data acquisition and processing}
        Hectospec offers 300 science fibres with 1.5\arcsec diameter which can be positioned over a circular area with a diameter of 1\degr. We used the grism 270 with 6\AA~resolution (1.2\AA~pix$^{-1}$) and wavelength coverage from 3650 to 9200\AA~centred at 5770\AA. The minimum separation allowed between the fibres is 20\arcsec, which forces our survey into more uniform sky coverage, rather than following the centrally concentrated cluster galaxy distribution.
        Two fibre configurations were observed with the MMT in service mode on 19 and 22 February 2014. Three consecutive 1200s exposures were acquired in each configuration for a total integration time of 1h per galaxy. The seeing was between 1\arcsec and 1.2\arcsec. No filters were used. From the 300 fibres available, 32 and 31 were used for sky emission in the first and second observing night, respectively. There were 6 unused fibres in each of the two configurations. Spectrophotometric calibration star observations were done during the second observing night.
        
        The data processing was done in IRAF\footnote{IRAF is distributed by the National Optical Astronomy Observatory, which is operated by the Association of Universities for Research in Astronomy (AURA) under cooperative agreement with the National Science Foundation.} with the occasional use of our own Python procedures. We did not use any dedicated Hectospec pipeline but we closely followed the ideology of SPECROAD pipeline \citep{Mink2007}.
        Following a correction for the time-constant problem of the Hectospec amplifiers, the processing was done with one bias frame for the whole observing run with no dark current image subtraction. The stability of  Hectospec allowed us to apply a master dome flat field image to correct for pixel-to-pixel sensitivity variations as well as the fringing pattern noticeable redward of 6500\AA. As with the bias, this approach was based on visual inspection of the available calibration frames. A correction for the cosmic ray events was made based on the variance between the median and every individual 20 minute exposure. The cut-off value was chosen such that effects on the data were minimised while still excluding the bulk of the cosmic rays. To fully remove the effects from the cosmic rays and the artefacts introduced by our cleaning algorithm we used a median combination to produce the final images. We also subtracted a 2D model of the scattered light.
        The correction for the different throughput of the individual fibres is achieved in two steps. First, wavelength dependent throughput correction is calculated from the sky flat-field images. The extracted 1D twilight sky spectra were wavelength calibrated. Following the SPECROAD pipeline approach, the spectra from apertures between 141 and 146 were averaged and then divided into all 300 twilight spectra. The so-produced ratio was fitted with a low-order function to produce a map of the wavelength-dependent part of the relative fibre throughput. The second step of the relative throughput correction was based on the flux measured in seven of the strongest sky emission lines which can reliably be measured. The relative scales were calculated for each fibre, based on the median of the line flux ratios excluding the two extremes. Unlike the first part of the throughput correction, this part does not have colour dependence. This is the final scaling applied to the individual fibres. The sky spectrum subtracted from each fibre was not additionally scaled as this scaling would be biased due to the signal from our targets.
        The area on the sky accepted by the individual fibres changes with the distance of the fibre from the centre of the focal plane. All the fibres were brought to the same sky coverage of 1.62 arcsec$^2$.
        The HeNeAr-lamp-based wavelength calibration achieved the expected accuracy for the instrument \citep{Fabricant2005, Fabricant2008}. 
        
        We examined all the sky spectra and removed the ones deviating from the mean leaving 24 and 20 \textit{clean} sky fibres from the two observing nights, respectively. A sky spectrum, averaged from the nearest six sky fibres, was subtracted from each target spectrum, as proposed by \cite{Fabricant2005}. We adopted this approach due to the smaller residuals from sky lines in the red end of the spectra when compared to other methods of sky subtraction \cite[e.g.][]{Mink2007}.
        
        The Hectospec pipeline provides a model of the sky absorption features around 6870\AA~and 7600\AA, this was divided into all the sky subtracted spectra, after which they were split into separate 1D spectra.
        
        Many Hectospec fibres are affected by light leakage from a fibre positioning LED. This is spread redward of 8500\AA. In the fashion of the SPECROAD pipeline, the continuum redward of 8500\AA~was fitted and the fit was subtracted only from the spectra showing a rising continuum in that region.
        
        The spectro-photometric star HZ44 was observed during the second observing night. We obtained the flux of the star from the CALSPEC Calibration Database and used it to calculate the transmission curve which, after median filtering, was applied to all the target spectra to achieve absolute flux calibration.
                
\subsubsection{Redshift estimation}
        We estimated the redshift of the targeted galaxies in two different ways. First, for the ones showing recognisable emission or absorption lines we used IRAF to fit a Gaussian to those lines. The lines used were {[\sc O\,i\,i]} 3727\AA, H$\alpha$ 6563\AA, H$\beta$ 4861.33\AA, {[\sc O\,i\,i\,i]} 5006.84\AA~and {[\sc O\,i\,i\,i]} 4958.91\AA~in emission and the calcium K and H+H$\epsilon$ lines in absorption at 3934\AA~and 3968\AA~together with Na 5892.50\AA. A visual quality flag was assigned to each redshift depending on the number, nature and strength of the lines used. This yielded 168 redshifts, based on at least two lines, designated $z_l$.
        A second estimation of the redshift was done with cross-correlation of the observed spectra with spectral templates from SDSS DR7\footnote{http://classic.sdss.org/dr7/algorithms/spectemplates/} \citep{SDSS1}. These redshifts are designated $z_{cc}$. The cross-corelation was done in IRAF, over the region from 3800 to 8000 \AA. The Fourier filter was set up following prescriptions from \citet{TonryDavis79} and \citet{Alpaslan2009}. The signal-to-noise ratio of the correlation peak is described by the Tonry and Davis' $r$ parameter \citep{TonryDavis79}. The cut-off value in $TD_r<$ 5.1, adopted in this work, was determined by examining the difference $|z_l - z_{cc}|$ as a function of $TD_r$ with a test similar to the one used by \citet{KurtzMink1998}. We followed this step with another visual inspection of all the spectra confirming the redshift of 190 more galaxies. We also add to our final $z$ catalogue the 49 targets with $z_{cc}$ and $TD_r >$ 5.1 for which we could not identify obvious emission/absorption features in their spectra. Those are referred to as $z_r$. 
        
        With this approach we are confident in the correct estimation of the redshifts for 407 (78\%) of our target galaxies. Those are distributed as follows: 168 (41\%) $z_l$, 190 (47\%) $z_{cc}$ and 49 (12\%) $z_r$.         The measured redshifts were corrected for the orbital motion of Earth.
\subsubsection{Uncertainties}
\begin{figure}
\centering
\includegraphics[width=\hsize]{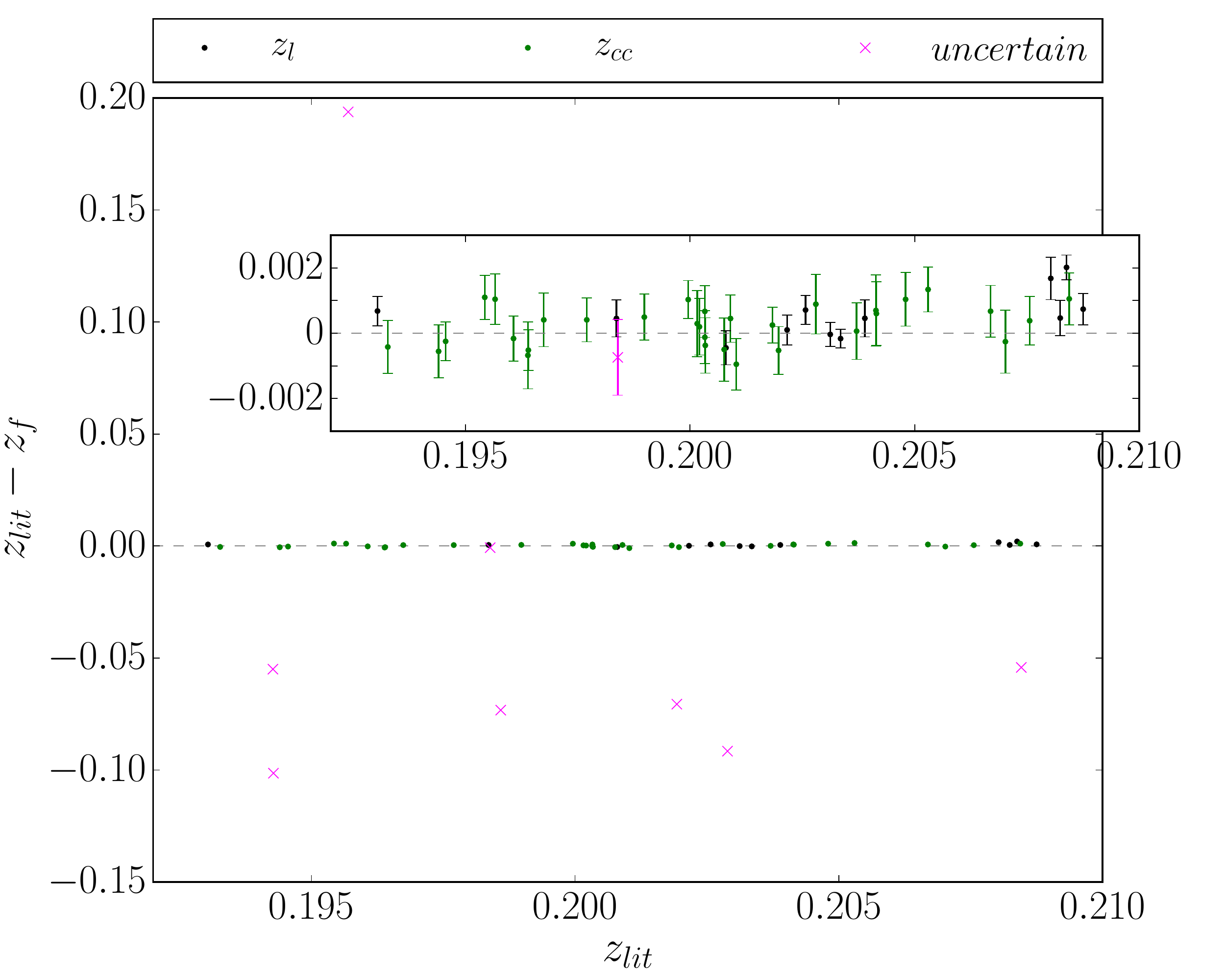}
  \caption{Comparison of our redshift estimations ($z_f$) with the published ones from G08 ($z_{lit}$) for the 52 common targets. The dots of different colours represent our secure estimations, while the magenta crosses are the uncertain ones. The inset is a zoom around the line of equality. The inset does not block any points. See the text for explanation of the separate $z_f$ estimations.}
  \label{check_zf}
\end{figure}

        Our final $z$ catalogue contains the $z_{cc}$ estimation of the redshift for 239 galaxies with uncertainty calculated as:
\begin{equation}
\delta z = \frac{3}{8} \frac{w}{(1+TD_r)},
\end{equation}
where $w$ is the width of the correlation peak. An interested reader can find an exhaustive discussion on the errors estimated by the IRAF task \textit{xcsao} in \cite{KurtzMink1998}.
        The final estimation $z_f$ for the 168 galaxies for which we also have $z_l$ is based on a weighted mean of the $z_l$ and $z_{cc}$, with the weights of $z_l$ based on the discrepancy between the $z$ shown by the different lines.
        The median error of all the galaxies in our catalogue is 78 km s$^{-1}$. Fig.~\ref{check_zf} shows a comparison between our measured redshifts and the ones published by G08 for the 52 common targets. The median error of \textit{cz} in their combined catalogue is 112 km s$^{-1}$. The standard deviation from the line of equality on Fig.~\ref{check_zf} (195 km s$^{-1}$) shows that we are likely underestimating the uncertainties of our measurements by ${\sim} 30 \%$. This is broadly consistent with the findings of \cite{KurtzMink1998} for the error estimations made with \textit{xcsao}. We increase the uncertainties in the final catalogue to compensate for this.
        Few things are apparent in Fig.~\ref{check_zf}. Both the cross-correlation and the line fitting methods have comparable spread around the zero difference line (green and black dots, respectively). Also, some of the "uncertain" redshift estimates ($TD_r<$ 5.1, magenta crosses on Fig.~\ref{check_zf}) are actually accurate to within the errors of the measurement. For 44 of the 52 targets in common with G08 we have enough signal in our MMT data to confirm their \textit{z} estimation with high certainty. For one of the 8 remaining targets we have a $z_{cc}$ estimation in agreement with the G08's value but with $TD_r<$ 5.1. We could not estimate the redshift of the remaining 7 targets.
\subsubsection{Completeness of the survey}\label{completeness}
\begin{figure*}
\centering
\includegraphics[width=\hsize]{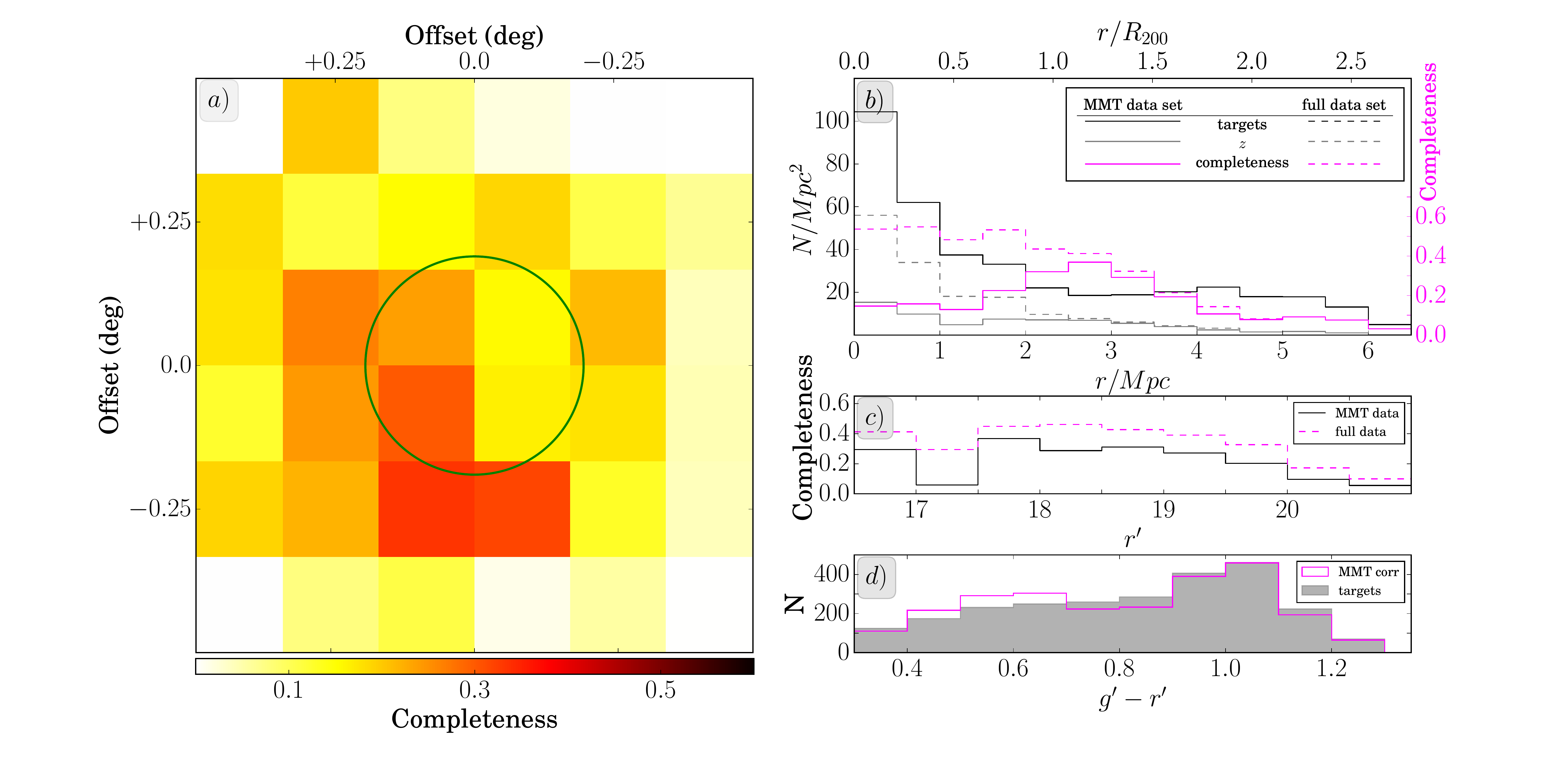}
  \caption{The completeness of our survey defined as the fraction of all possible targets for which $z$ was successfully estimated. Panel \textit{a} shows the 2D distribution of the completeness on the sky for the MMT data set. The green circle has a radius equal to R$_\mathrm{200}$ of the cluster (derived in Section~\ref{glob_params}). Panel \textit{b} shows the completeness, and number density distribution, of the MMT survey and the full data set as a function of cluster-centric distance. Panel \textit{c} shows the completeness as a function of $r'$ magnitude. Panel \textit{d} shows the effect of the completeness correction calculated from panel \textit{c}. The filled, grey histogram shows the $g'-r'$ colour distribution of all the potential targets. The solid magenta line shows the colour distribution of the galaxies with $z$ estimate from the MMT data set corrected for completeness.}
  \label{2d_compl}
\end{figure*}
        We adopted an approach similar to \citet{Milvang2008} in estimating the completeness of the survey. Completeness is defined as the ratio of the number of galaxies with successful $z$ estimate to the total number of potential targets. The potential targets are all the extended objects in our photometric catalogue, within the confinements of the spectroscopic survey that satisfy our magnitude and colour criteria (see Section~\ref{target_selection}).
        
        Fig.~\ref{2d_compl} shows the distribution of completeness on the sky and as a function of magnitude and cluster-centric distance. Panel \textit{a} shows the completeness in bins of R.A. and Dec for the MMT data set. The green circle has a radius equal to R$_\mathrm{200}$ of A520 (derived in Section~\ref{glob_params}). This completeness is used to correct the sky distribution of various cluster members shown on Fig.~\ref{proj_density}. Panel \textit{b} shows the completeness of the MMT survey (solid, magenta line), and the full data set (dashed, magenta line) as a function of cluster-centric distance. This completeness is very important for the estimation of global cluster parameters which can be overestimated from multi-fibre spectroscopic surveys which tend to under-sample the inner, denser parts of the clusters and over-sample the outskirts increasing the number of interlopers \citep{Biviano2006}. As shown on panel \textit{b}, the densest parts of A520 are indeed undersampled by the MMT survey. However, the full data set, which we use to determine cluster membership and global cluster parameters in Section~\ref{glob_params}, does show a relatively flat sampling rate of the galaxy distribution from the centre out to ${\sim}3$ Mpc, or $\sim1.5$R$_{200}$, at approximately $50\%$. Because the MMT survey used only two fibre configurations with a preference of targets with cluster-centric distances smaller than 3Mpc, and because of the technical limitations for fibre positioning, it has an irregular radial completeness, which peaks at ${\sim} 50\%$ around $3$ Mpc but drops towards the centre of the cluster, demonstrating the need for sampling the cluster with multiple fibre configurations. The black and grey lines on panel \textit{b} show the number density distribution of targets and galaxies with successful $z$ estimate, respectively.
        
                Panel \textit{c} shows the completeness of the MMT data set (black, solid line) and the full data set (magenta, dashed line) as a function of $r'$ magnitude. The mean completeness of our MMT survey is 16\% (407/2523) over the entire magnitude range. Due to fibre positioning limitations, in the magnitude range $17 < r' < 17.5$, we have observed only 1 of the 17 available targets. The completeness drops significantly for targets with $r'>$19. This completeness curve is used to correct the distribution of star forming and blue galaxies shown in Figs.~\ref{frac_rad_substr} and \ref{em_line_frac}.
                
                To demonstrate the effectiveness of the completeness correction on the last panel \textit{d} of Fig.~\ref{2d_compl} we plot the $g'-r'$ colour distribution of all potential targets (grey filled histogram) and of all the galaxies from the MMT data set after applying the magnitude-based completeness correction (solid, magenta line). Even though the target selection doesn't give preference to galaxies of any particular colour, the limitations due to fibre collisions, the drop in completeness with absolute magnitude, and the fact that we only use two fibre configurations could potentially introduce a colour bias in the final data set. To test this, we run a two-sample K-S test on the colour distribution of all the potential targets (grey filled histogram in Fig.\ref{2d_compl}\textit{d}) and all the galaxies from the MMT data set, before and after applying the correction for completeness. Indeed, before the correction is applied, the test gives a rather low value of $p^{K-S}_{uncorr}$=0.05, which indicates potentially different colour distributions. After applying the magnitude-based completeness correction, the solid magenta line in Fig.~\ref{2d_compl} suggests that blue galaxies are slightly over-represented in the MMT data set. However, the K-S test shows that the two samples in fact have near identical colour distributions with $p^{K-S}_{corr}$=0.97, and could be drawn from the same parent distribution.
        
\begin{figure}
\centering
\includegraphics[width=\hsize]{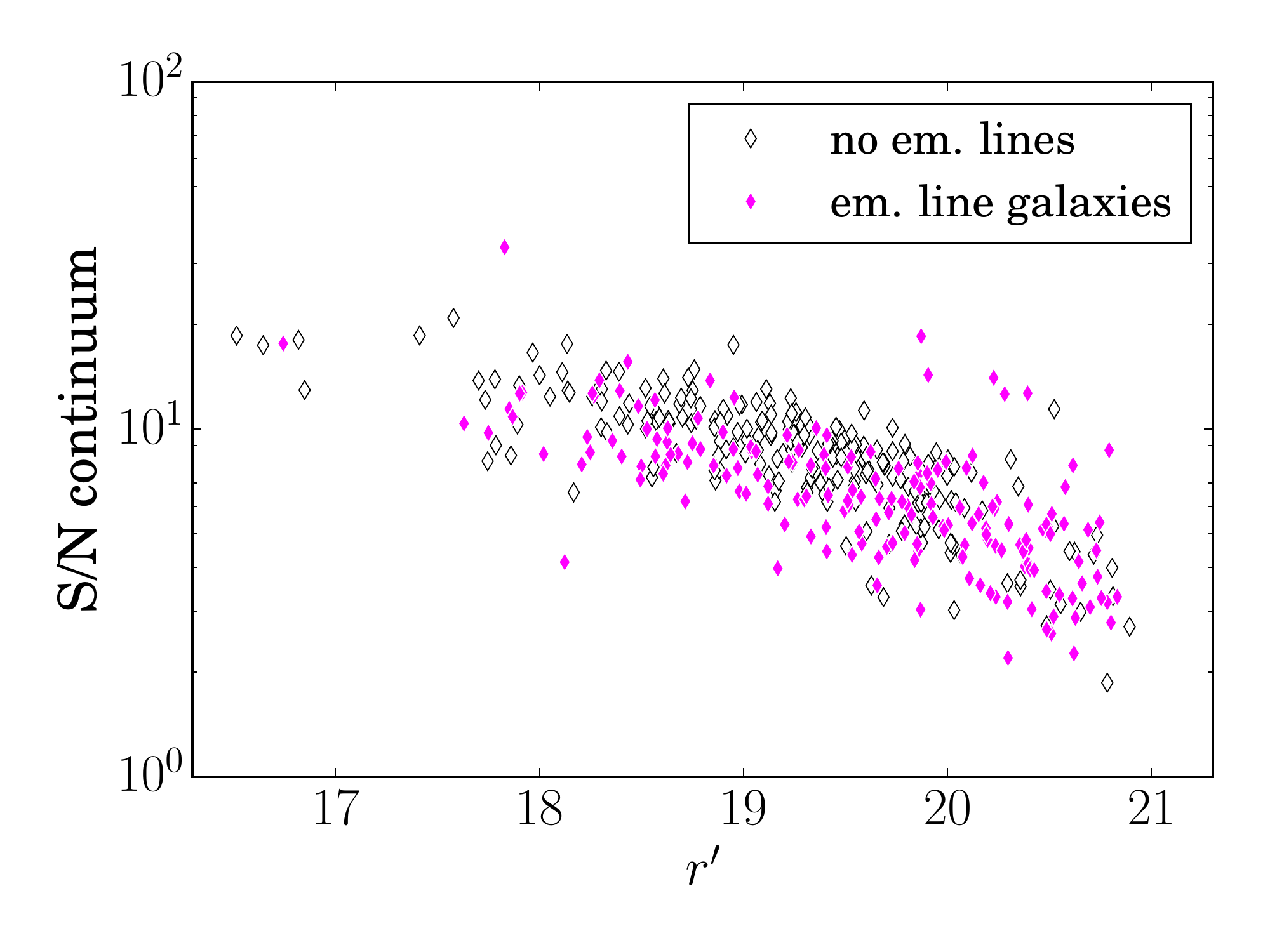}
  \caption{The signal to noise ratio in the continuum versus integral $r'$ magnitude for all the galaxies in our final redshift catalogue. The open black diamonds denote galaxies without emission lines and the filled magenta ones - galaxies whose redshift was determined from emission lines.}  \label{sn_mag}
\end{figure}

        Because it is easier to measure redshifts from emission lines it is possible to over-represent emission-line galaxies in the final catalogue. Fig.~\ref{sn_mag} tests for the presence of such a bias in a way presented in \citet{Verdugo2008}. It shows the signal-to-noise ratio ($S/N$) in the continuum as a function of the galaxy total magnitude, separating the galaxies with and without emission lines. The continuum is fitted iteratively with a fourth-order polynomial, clipping at $4\sigma$ on each of the five iterations. The final distribution of galaxies in the so-defined plane is very uniform without a significant difference in the distribution of emission and absorption line objects, demonstrating that even at the faint end of our survey we regularly estimate redshifts from absorption lines.
\subsection{Comparison sample}\label{comp_sample}
        As a comparison sample we use ten clusters from the ACReS survey \citep[Arizona Cluster Redshift Survey,][]{Haines2013}. This survey combines spectroscopic data of galaxies in 30 clusters with $0.15<z<0.3$, drawn from the LoCuSS \citep{Smith2010a} sample. ACReS uses the same instrument (Hectospec), and the same instrument setup as our MMT survey of A520.

                From the 30 available clusters we select only those in the redshift range $0.15<z<0.25$. We also remove known cluster mergers (e.g. A115, A1914), as well as the ones that show obvious multi-component structure, by visually inspecting position-velocity plots of all the cluster members. We also remove the ones that are not covered by SDSS, as we used the SDSS $g'$ and $r'$ photometry to calculate stellar masses, and to correct for completeness. This leaves a sample of 10 clusters with a total of 3218 spectroscopically confirmed member galaxies. The membership of the comparison clusters and their stellar masses are determined in the same way as for A520 (see Sections~\ref{glob_params} and \ref{st_mass}).

        ACReS target selection is based on J and K bands near-infrared imaging done with the United Kingdom Infrared Telescope and Mayall telescope at Kitt Peak National Observatory. In contrast, A520 target selection is based on optical $g'$ and $r'$ band imaging. Thanks to its target selection, which very efficiently removes foreground and background objects, ACReS is highly complete for cluster-member galaxies. This is shown in \citet{Haines2013}. In order to make the two samples compatible we apply our colour and magnitude selection to the ACReS data discarding all the galaxies that would not have been selected for observation in A520. This removes $\sim10\%$ of the initial number of galaxies with $z$ from the ACReS clusters. After that we estimate the completeness of ACReS in the same way we did for A520 in Section~\ref{completeness}, using all the extended objects in SDSS within the sky coverage of the spectroscopic survey and within the colour and magnitude limits presented in Section~\ref{target_selection}. We use the completeness curves calculated this way to correct the results presented in Figures~\ref{em_line_frac} and \ref{proj_density}.
                
        In addition to the MMT observations, ACReS includes a number of cluster member galaxies with previously published redshifts. We use this combined data set to calculate global cluster parameters and cluster membership of the comparison sample. Table~\ref{acres_tbl} lists the main properties of the comparison clusters and A520 related to the observations and determined in the following Section~\ref{results}. When estimating the current and past star formation properties of galaxies, in order to guarantee a fair comparison, we only use the MMT part of the ACReS data, similarly to the MMT data set for A520. This guarantees uniformity of the sample, and the removal of all observational differences usually associated with heterogeneous data sets.
\\
\begin{table*}
\centering
\caption{The comparison clusters and A520 with their main parameters. N$_\mathrm{m}$ is the number of cluster members from the ACReS survey (combined data set) or in the case of A520 from the MMT data set (full data set). N$_\mathrm{pass}$, N$_\mathrm{sf}$ and N$_\mathrm{rq}$ are the number(fraction) of passive, star forming and recently quenched galaxies, respectively. N$_\mathrm{AGN}$ is the number(fraction) of emission-line galaxies who's spectra are dominated by AGN emission. All the numbers and fractions are calculated among all the members of the given cluster.}
\label{acres_tbl}
\begin{tabular}{l l l r r r r r r r}
\hline
\hline
Cluster & $z$ & N$_\mathrm{m}$ & $\sigma_v$ & R$_\mathrm{200}$ & M$_\mathrm{200}$ & N$_\mathrm{pass}$ & N$_\mathrm{sf}$ & N$_\mathrm{rq}$ & N$_\mathrm{AGN}$\\
 & & & km s$^{-1}$ & Mpc & [10$^{14}$M$_{\odot}$] & (\%) & (\%) & (\%) & (\%)\\
\hline
\object{A665} & 0.1818 & 310(360) & 1268$^{+103}_{-109}$ & 2.87$^{+0.08}_{-0.44}$ & 21.5$^{+1.5}_{-6.8}$ & 205(66) & 100(32) & 14(4) & 5(5) \\
\object{A1689} & 0.1832 & 519(934) & 1779$^{+124}_{-117}$ & 4.02$^{+0.21}_{-0.35}$ & 59.2$^{+6.4}_{-9.1}$ & 308(59) & 185(35) & 24(4) & 26(12) \\
\object{A383} & 0.1883 & 164(260) & 898$^{+84}_{-95}$ & 2.02$^{+0.12}_{-0.28}$ & 7.6$^{+1.4}_{-2.5}$ & 88(53) & 63(38) & 23(14) & 13(17) \\
\object{A291} & 0.196 & 131(131) & 682$^{+123}_{-134}$ & 1.53$^{+0.50}_{-0.02}$ & 3.3$^{+2.8}_{-0.1}$ & 79(60) & 46(35) & 0(0) & 6(12) \\
\object{A963} & 0.205 & 414(492) & 1199$^{+112}_{-112}$ & 2.68$^{+0.37}_{-0.14}$ & 17.9$^{+7.5}_{-2.4}$ & 206(49) & 199(48) & 21(5) & 9(4) \\
\object{Z1693}\tablefootmark{a} & 0.2248 & 212(265) & 683$^{+56}_{-52}$ & 1.51$^{+0.09}_{-0.14}$ & 3.3$^{+0.7}_{-1.0}$ & 129(60) & 74(34) & 16(7) & 9(11) \\
\object{A1763} & 0.2279 & 305(447) & 1399$^{+122}_{-134}$ & 3.09$^{+0.37}_{-0.21}$ & 28.1$^{+9.1}_{-4.3}$ & 156(51) & 122(40) & 38(12) & 27(18) \\\object{A2390} & 0.2329 & 356(552) & 1336$^{+112}_{-112}$ & 2.94$^{+0.20}_{-0.29}$ & 24.4$^{+5.3}_{-6.3}$ & 193(54) & 135(37) & 5(1) & 28(17) \\
\object{Z2089}\tablefootmark{b} & 0.2347 & 141(141) & 761$^{+76}_{-73}$ & 1.68$^{+0.24}_{-0.10}$ & 4.5$^{+2.1}_{-0.7}$ & 80(56) & 50(35) & 14(9) & 11(18) \\
\object{R2129}\tablefootmark{c} & 0.235 & 236(307) & 907$^{+76}_{-80}$ & 2.00$^{+0.13}_{-0.22}$ & 7.6$^{+1.4}_{-1.9}$ & 148(62) & 74(31) & 16(6) & 14(16) \\
\textbf{A520} & 0.2007 & 190(315) & 1036$^{+101}_{-97}$ & 2.32$^{+0.23}_{-0.22}$ & 11.6$^{+3.7}_{-3.0}$ & 137(72) & 50(26) & 12(6) & 3(5) \\
\hline
\end{tabular}
\tablefoot{
\tablefoottext{a}{\object{ZwCl0823.2+0425};}
\tablefoottext{b}{\object{ZwCl0857.9+2107};}
\tablefoottext{c}{\object{RXJ2129.6+0005/RBS1748};}}
\end{table*}
\subsection{Stellar mass estimation}\label{st_mass}
         We calculate the stellar masses of the galaxies in A520 and the comparison clusters using colour dependent mass to light ratios $\Upsilon_r$ calculated according to an empirical recipe by \citet{Zibetti2009} using Chabrier IMF, and the absolute $r'$ magnitudes and K-corrected $g'-r'$ colours derived in Section~\ref{add_imaging}. We adopt a solar absolute $r'$-band magnitude of 4.65.

\section{Galaxy classification}\label{gal_class}
We separate galaxies into blue/red if their integral $g'-r'$ colours are lower/greater than or equal to 0.9.

\subsection{Star formation and AGN diagnosis}
        The emission lines observed in the spectra of some galaxies come from the excited gas in the inter-stellar medium (ISM). The excitation usually comes from sources of ionising radiation, active galactic nucleus (AGN) or young massive stars, which in itself is indicative of ongoing star formation \citep{BinneyMerrifield98}. The radiation from the two have different characteristics and can be diagnosed by comparing the relative intensities of specific emission lines \citep{BPT1981}.

        We diagnose sources with AGN contribution among the emission line galaxies by using three separate optical diagnostic plots. One from \citet{Lamareille2010}, known as the "blue diagram", which compares the ratio of {[\sc O\,i\,i\,i]}/$H\beta$ with that of {[\sc O\,i\,i]}/$H\beta$ and minimises contamination according to their Eq.1. The other two diagnostics plots are from \citet{BPT1981}, known as the "BPT diagram". We use two separate versions of the latter, one using the ratio of {[\sc N\,i\,i]}/$H\alpha$ and the other {[\sc S\,i\,i]}/$H\alpha$, versus {[\sc O\,i\,i\,i]}/$H\beta$ with demarcation lines from \citet{Kewley2006}. We adopted this approach to maximise the number of galaxies these diagnostics can be applied to. If the given spectrum is identified as AGN dominated by any of the tests, we label the galaxy as AGN and remove it from the sample of star forming galaxies.
        
        Approximately one quarter of all the galaxies in the MMT survey have a measurable flux in the needed emission lines for at least one of the diagnostics. In total we find 13 galaxies whose spectra are likely dominated by AGN, 3 of which are also classified as members of A520.
        
        The same process was repeated with the comparison sample. The median number of AGNs per cluster in the comparison sample is 13.5, and the median fraction is 15\%. This compares well with the $\sim$20\% fraction found in the local universe \citep{Miller2003}, and $\sim$17\% found at z=0.4 by \citet{Sobral2016}. As indicated on Table~\ref{acres_tbl}, A520 is at the lowest extreme of the distribution of AGN fractions, together with A963 and A665.

        If not coming from an AGN, the presence of emission lines in the spectrum of a given galaxy is a sign of ongoing star formation. We split our galaxies into star forming or passive according to the presence/absence of {[\sc O\,i\,i]} and $H\alpha$ emission lines in their spectra. If any of these lines have equivalent width EW $\geq 3 \AA$, and the spectrum is not dominated by AGN, we consider the galaxy to be star forming.

\subsection{Recently quenched galaxies}\label{rec_q}
\begin{figure}
\centering
\includegraphics[width=\hsize]{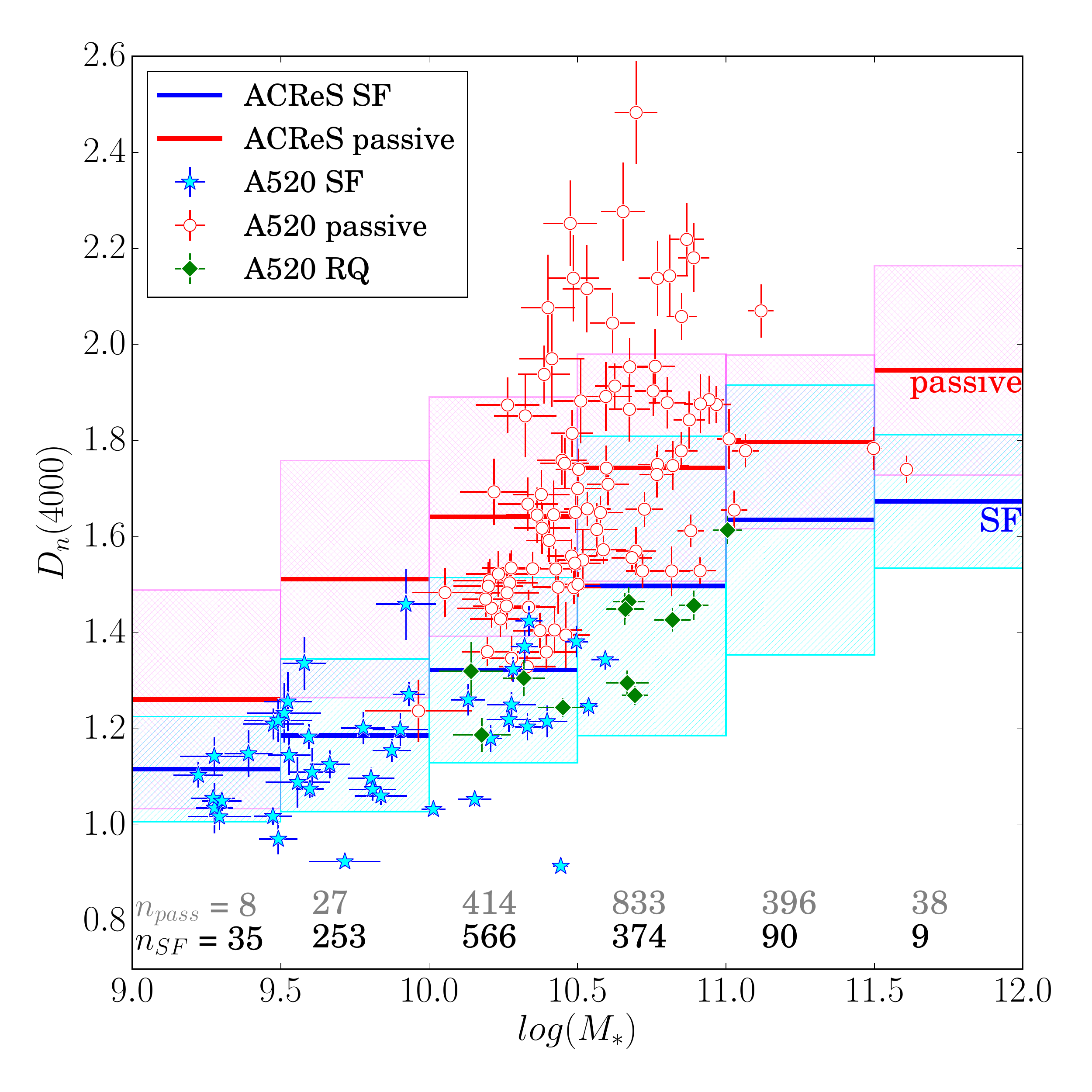}
  \caption{The strength of the 4000\AA~break as a function of stellar mass. The mean $D_n(4000)$ value for the star forming galaxies in the comparison clusters, in bins of stellar mass, is shown with the thick blue lines. The thick red lines show the same for the passive members of the comparison clusters. The shaded regions in cyan and magenta show the standard deviation around the mean values for the star forming and passive galaxies, respectively. The A520 galaxies are split into three groups. The star forming ones are shown with blue stars, the passive ones with empty red circles and the recently quenched ones with green diamonds. The last group contains the passive galaxies with $D_n(4000)$ lower than the average of the star forming ACReS galaxies with the same stellar mass. The number of passive and star forming galaxies in the comparison clusters is indicated in each stellar-mass bin.}
  \label{d4_stmass}
\end{figure}
\begin{figure}
\centering
\includegraphics[width=\hsize]{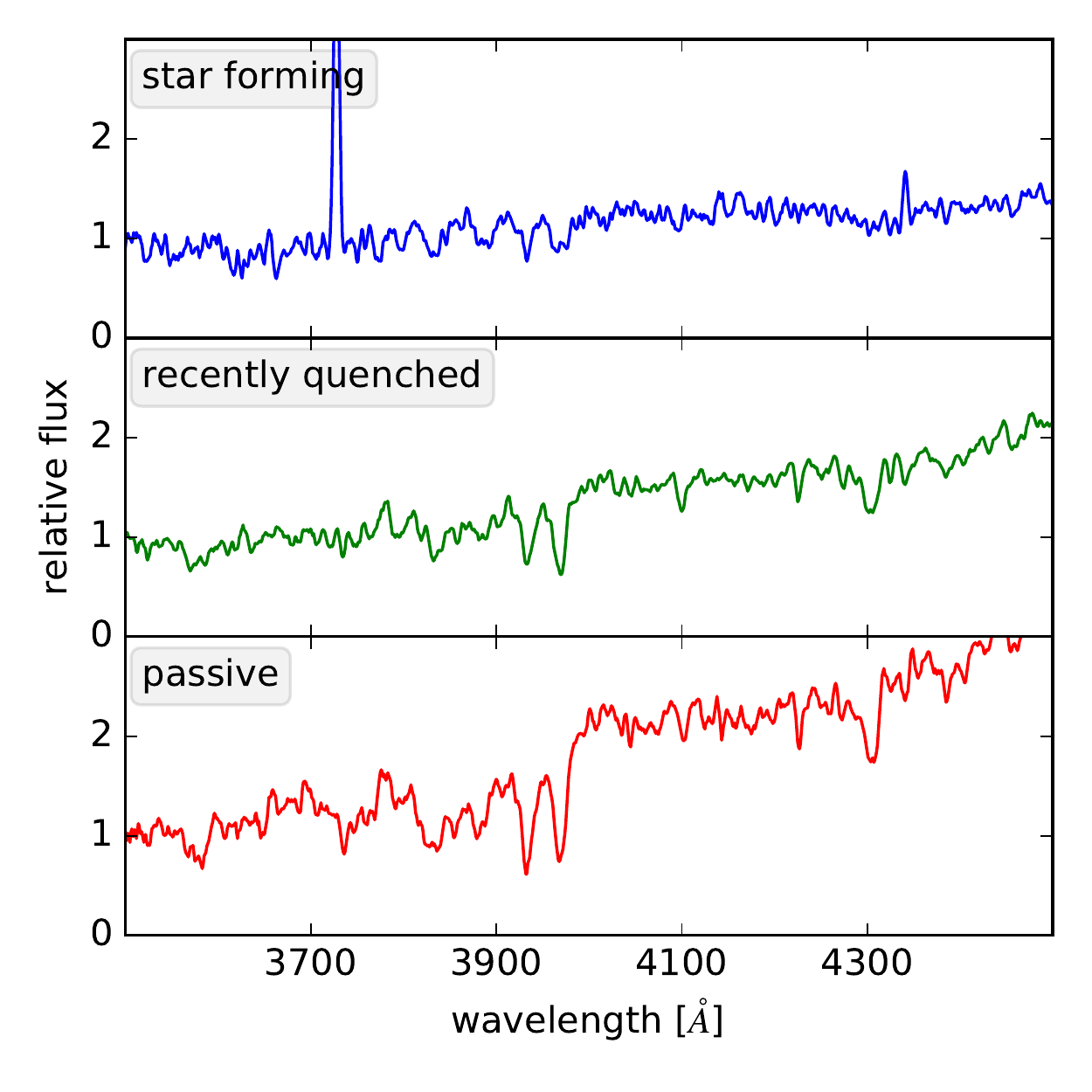}
  \caption{Example of the three spectroscopic types of galaxies. The middle panel shows the sum of the spectra of the 12 recently quenched galaxies. The top and bottom panels show the sum of 12 spectra randomly chosen from the star forming and passive galaxies, respectively.}
  \label{all_sp}
\end{figure}

        We use the strength of the 4000\AA~break - $D_n(4000)$ as an indicator of the time passed since the most recent star formation event. The 4000\AA~break constitutes a drop in the strength of the continuum blueward of 4000\AA~caused by the cumulative effects of a series of metal absorption lines in the atmospheres of low-mass stars. Here we adopt the "narrow" definition of the 4000\AA~break- $D_n(4000)$ \citep{Balogh1999}, which is the ratio of the flux between 4000\AA~and 4100\AA~and between 3850\AA~and 3950\AA, for its lower sensitivity to reddening. The $D_n(4000)$ index is often used to quantify the luminosity-weighted age of the stars in a galaxy as it increases with the time since the end of the star formation event \citep{DresslerShectman1987}. Before measuring $D_n(4000)$ we smooth all the MMT spectra with a Gaussian with $\sigma = 2$ pixels to increase the S/N. We note that the raw MMT spectra are sampled at 5 pixels per resolution element. 78\% (149/190) of the cluster members' spectra have sufficient
S/N for reliable measurement of $D_n(4000)$. These are plotted in Fig.\ref{d4_stmass}.
        
        The age and metallicity of the stellar population (and hence the $D_n(4000)$ index) are known to show dependence on the stellar mass of the galaxy \citep[see e.g.][]{Geller2014, Geller2016, Haines2016}. In order to remove this dependence and select recently quenched galaxies we plot in Fig.~\ref{d4_stmass} $D_n(4000)$ as a function of the galaxy stellar mass. The galaxies in the comparison clusters are represented by the mean $D_n(4000)$ in bins of stellar mass. For star forming galaxies this mean is shown by the thick blue line and its standard deviation with the shaded cyan region. The mean $D_n(4000)$ of the passive galaxies is shown with the thick red line, and its standard deviation with the magenta shaded region. The A520 members are also split into star forming and passive, shown with blue stars and red circles, respectively. At given stellar mass, the passive A520 members that have $D_n(4000)$ values lower than the average of the star forming comparison galaxies are classified as recently quenched galaxies. These are shown with green diamonds and indicated with RQ in figure legends. These galaxies have no emission lines but have low $D_n(4000)$ indicating the presence of a young stellar population. The numbers at the bottom of Fig.~\ref{d4_stmass} show the number of passive and star forming galaxies in each bin of stellar mass in the comparison clusters.
        
        In order to estimate the lifetime of the recently quenched phase we used the \cite{Bruzual&Charlot2003} model spectra library to observe the evolution of the $D_n(4000)$ index in galaxies with complex star formation histories. We used star bursts overlaid on top of old stellar population, and producing up to 12\% of the final stellar mass of the galaxy. These models used initial mass function by \cite{Kroupa2001}. All the tests showed that a recent star burst would leave a detectable trace in the $D_n(4000)$ index for $\lesssim$0.5~Gyr.
        
        On Fig. \ref{all_sp} we show one spectrum, representing each spectral class. For the recently quenched galaxies (middle panel), the plotted spectrum is a sum of the spectra of all 12 such galaxies found in A520. The top and bottom panels show a sum of 12 spectra randomly chosen from the star forming and passive A520 members, respectively. The spectrum of the star forming galaxies has a strong {[\sc O\,i\,i]} 3727\AA~line in emission. Although both the middle and bottom panels show the 4000\AA~break, it is noticeably stronger in the bottom one. Also noticeable is the different relative depth of the H+H$\epsilon$ and K absorption lines of ionised calcium, which is also indicative of the age of the stellar populations.
        
        One caveat, that has to be kept in mind when inferring galaxy properties on the basis of fibre spectroscopy, is that this kind of analysis is dominated by the central parts of the galaxies, even at the redshifts discussed in this article. The diameter of the Hectospec fibres is 1.5\arcsec which at the redshift of A520 means we are gathering light only within a radius of ${\sim}2.5$ kpc from the centre. This region, residing in the deepest potential well of the galaxy halo, likely experiences the environmental effects differently from the outskirts of the galaxy. Section~\ref{discussion} contains a more detailed discussion on how this caveat affects the conclusions drawn from the results presented in the following section.

\section{Results}\label{results}
\begin{figure}
\centering
\includegraphics[width=\hsize]{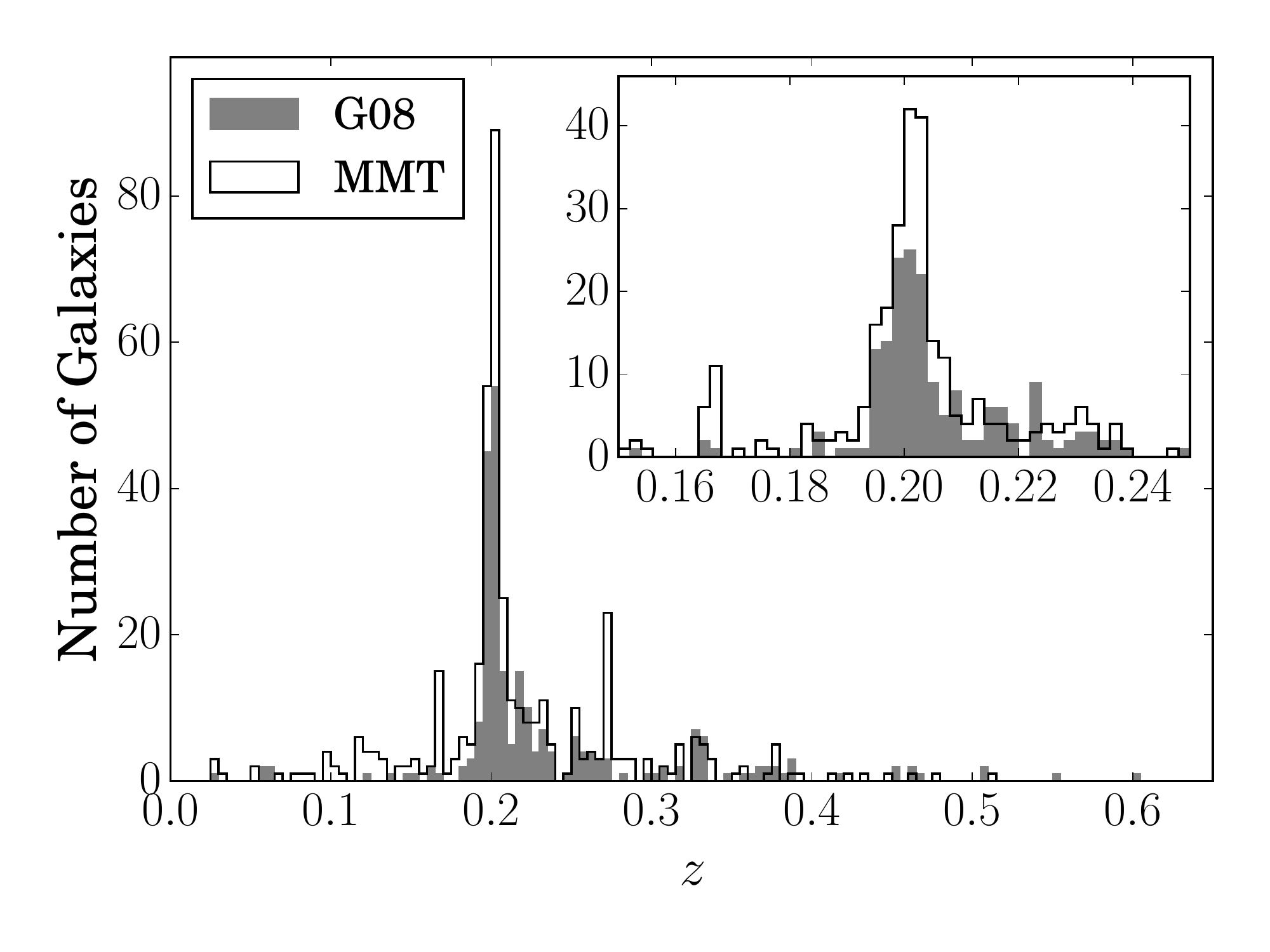}
  \caption{The redshift distribution of all the galaxies with redshift measurements from our MMT survey (unfilled histogram), and the ones published by G08 (grey, filled histogram). The inset shows a close-up of the cluster region.}
  \label{z_distr}
\end{figure}

        Fig.~\ref{z_distr} shows the redshift distribution of the 407 galaxies with MMT redshift estimates (unfilled black histogram) together with the $z$ distribution of the 293 galaxies published by G08 (filled grey histogram). The peak around $z=0.2$ corresponds to A520. We note that because the merger happens very close to the plane of the sky (G08) the two main galaxy concentrations have very small separation in velocity and are indistinguishable on this plot.

\subsection{Global cluster parameters and membership}\label{glob_params}
\begin{figure}
\centering
\includegraphics[width=\hsize]{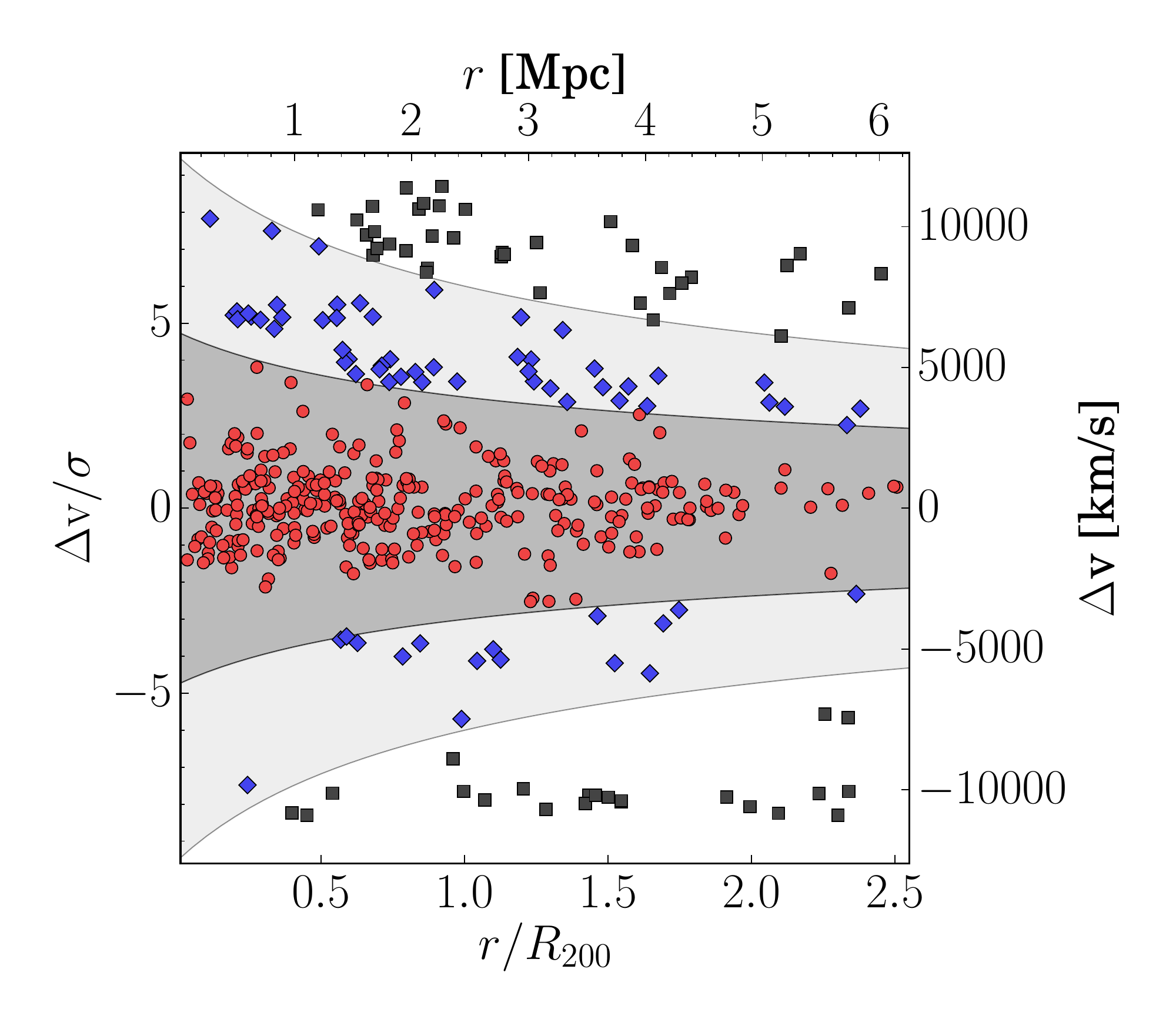}
  \caption{A projected phase-space diagram of A520. The galaxies from the full data set are plotted as a function of the cluster-centric distance (normalised by the cluster's R$_\mathrm{200}$) and peculiar velocity (normalised by the velocity dispersion of the cluster). The cluster members, according to the SIS analysis (see text for details) are plotted with red circles. Galaxies in the near field are shown with blue diamonds and non members with black squares. The curves outlining the regions populated by cluster members and near field galaxies are drawn according to Eq.~\ref{sigmar} with normalisation of 3$\sigma$ and 6$\sigma$ at R$_\mathrm{200}$, respectively.}
  \label{SIS}
\end{figure}

        The analysis presented in this, and the following Section~\ref{substructure} is based on the full data set, unless stated otherwise.
        
        In order to determine cluster membership and global cluster parameters we use the full data sets for A520 and the comparison sample, and assume that the clusters are well represented by a Singular Isothermal Sphere \citep[SIS,][]{Carlberg97}. As a centre of A520 we choose the peak of the X-ray emission at  R.A.=$04^{\mathrm{h}}54^{\mathrm{m}}07\fs44$, Dec=$+02\degr 55\arcmin 12\farcs0$, as adopted by G08. The derived central redshift  of the cluster is $z=0.2007$ and velocity dispersion, corrected with $(1+z)^{-1}$, is $\sigma_v=1036^{+101}_{-97}$ km s$^{-1}$. The cited uncertainties represent $95\%$ confidence level from 1000 bootstrap realisations. The radius within which the mean density is 200 times the critical density of the universe, as defined in Eq.\ref{r200} taken from \citet{Carlberg97}, is R$_\mathrm{200}=2.32^{+0.23}_{-0.22}$Mpc and the mass within that radius M$_\mathrm{200}=11.6^{+3.7}_{-3.0}\times10^{14}$M$_{\odot}$.
        
        The galaxy membership selection is shown on Fig.~\ref{SIS} and is based on a method used by \citet{Pimbblet06}. The radial velocity dispersion is expressed in Eq.~\ref{sigmar}

\begin{minipage}{.3\linewidth}
\begin{equation}
\sigma^2(r) = \frac{B}{b+R}
\label{sigmar}
\end{equation}
\end{minipage}%
\begin{minipage}{.4\linewidth}
\begin{equation}
\mathrm{;\,\,\,\,\,\,\,\,\,\,}
  R_\mathrm{200} = \frac{\sqrt{3}\sigma}{10H(z)},\end{equation}

\end{minipage}
\newline

where R=r/R$_\mathrm{200}$. B=1/4 and b=0.66 are the parameters determined by \citet{Carlberg97} by fitting the number density profiles of an ensemble of clusters. The demarcation curves on Fig.~\ref{SIS} are drawn at 3$\sigma(r)$ and 6$\sigma(r)$ at R$_\mathrm{200}$ to separate the members, near field and field galaxies. The final number of A520 members is 190(315) from the MMT data set (full data set). The number of galaxies in the near field region is 33(66).

        We repeat the same procedure with all the clusters from the comparison sample, with the number of members indicated in Table~\ref{acres_tbl}. The centres of the clusters are taken from NED, except for R2129 which is taken from \citet{Quillen08}. 
        
        Assuming that a merging cluster of galaxies is well represented by SIS might seem unreasonable. However, the match between the calculated velocity dispersion 1144$_{-67}^{+64}$ km s$^{-1}$ \citep{Hoekstra2015}, and 1066$_{-61}^{+67}$ km s$^{-1}$ (G08, for the whole system), and total cluster mass (M$_{vir}^{NFW}$=15.3$\pm$3.0$\times10^{14}$~h$^{-1}_{70}$~M$_{\odot}$ \citep{Hoekstra2015} and M$_{vir}$=8$\pm$2$\times10^{14}$~h$^{-1}_{70}$~M$_{\odot}$ (G08)) with the values shown in Table~\ref{acres_tbl} is very encouraging, and shows that the SIS approximation and the related membership selection used in this work are reasonable. We note that the separation between the centres of the two main galaxy concentrations in A520 (P2 and P4 on Fig.~\ref{ds_res} and Fig.~\ref{proj_density}), projected on the sky, is 0.84~Mpc, which is significantly smaller than the estimated R$_\mathrm{200}$ = 2.32~Mpc. Also the velocity distribution of the galaxies populating these two regions are indistinguishable from each other, and the overall velocity distribution of all the cluster members could be well approximated with a Gaussian.
        
        The goal of this work is to compare A520 to other clusters at the same redshift that have not experienced recent major mergers. Since exactly the same approach was used to select cluster members and to calculate R$_\mathrm{200}$ for A520 and the comparison sample we are confident that the analysis presented in the following sections is not affected by the assumptions made here. As we see in Sections \ref{sf_gals} and \ref{gal_distr}, the main differences between A520 and the comparison clusters concern the central-most parts of the clusters which are the least affected by the adopted membership selection.
        
\subsection{Substructure}\label{substructure}
\begin{figure}
\centering
\includegraphics[width=\hsize]{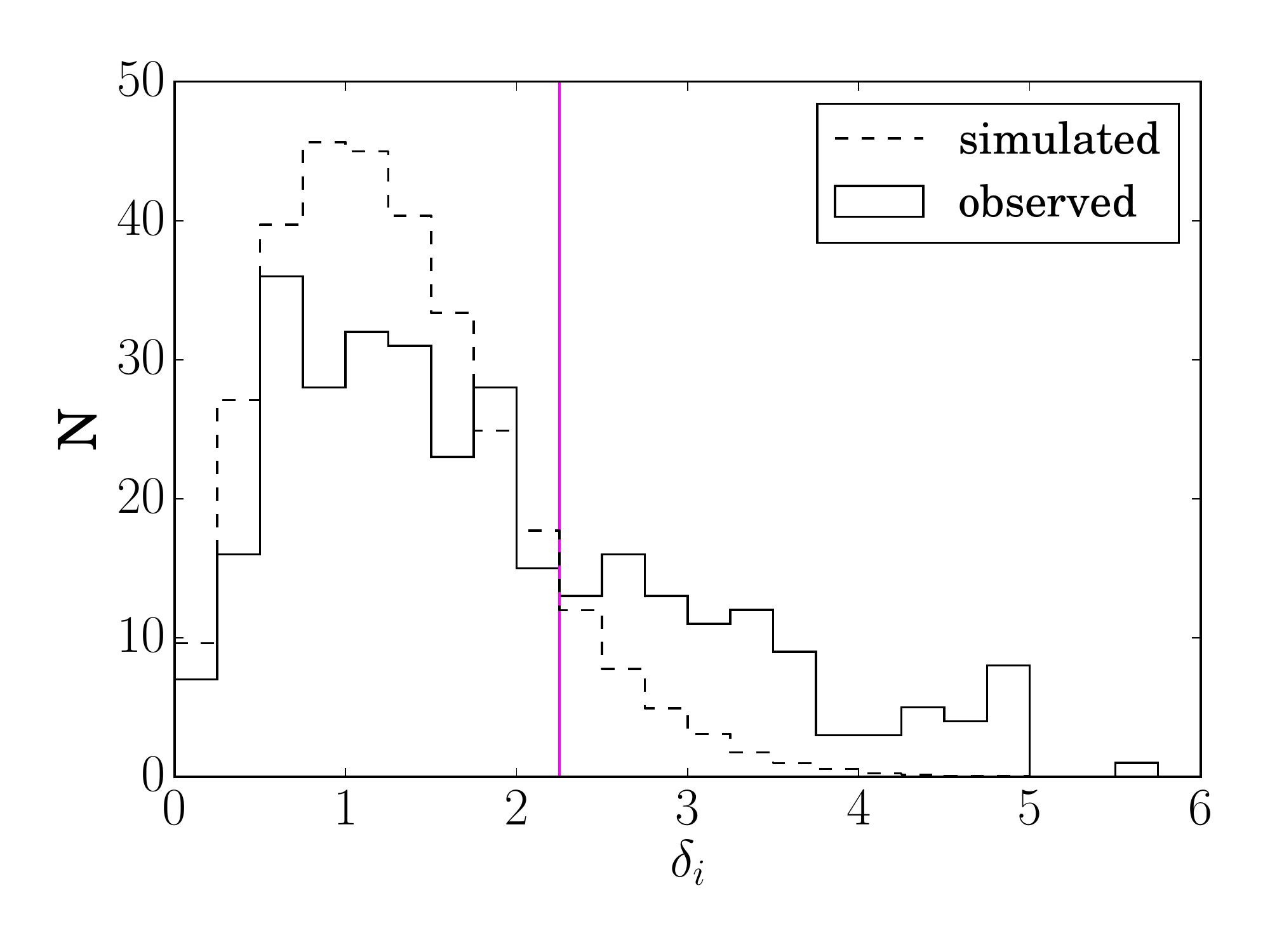}
  \caption{Distribution of the deviations among all the cluster members of A520 (solid line) and the members of 1000 mock clusters (dashed line, normalised). The cut-off value of $\delta_i$, shown with the solid magenta line, is determined from this plot as $\delta_i = 2.25$.}
  \label{dev_distr}
\end{figure}
\begin{figure}
\centering
\includegraphics[width=\hsize]{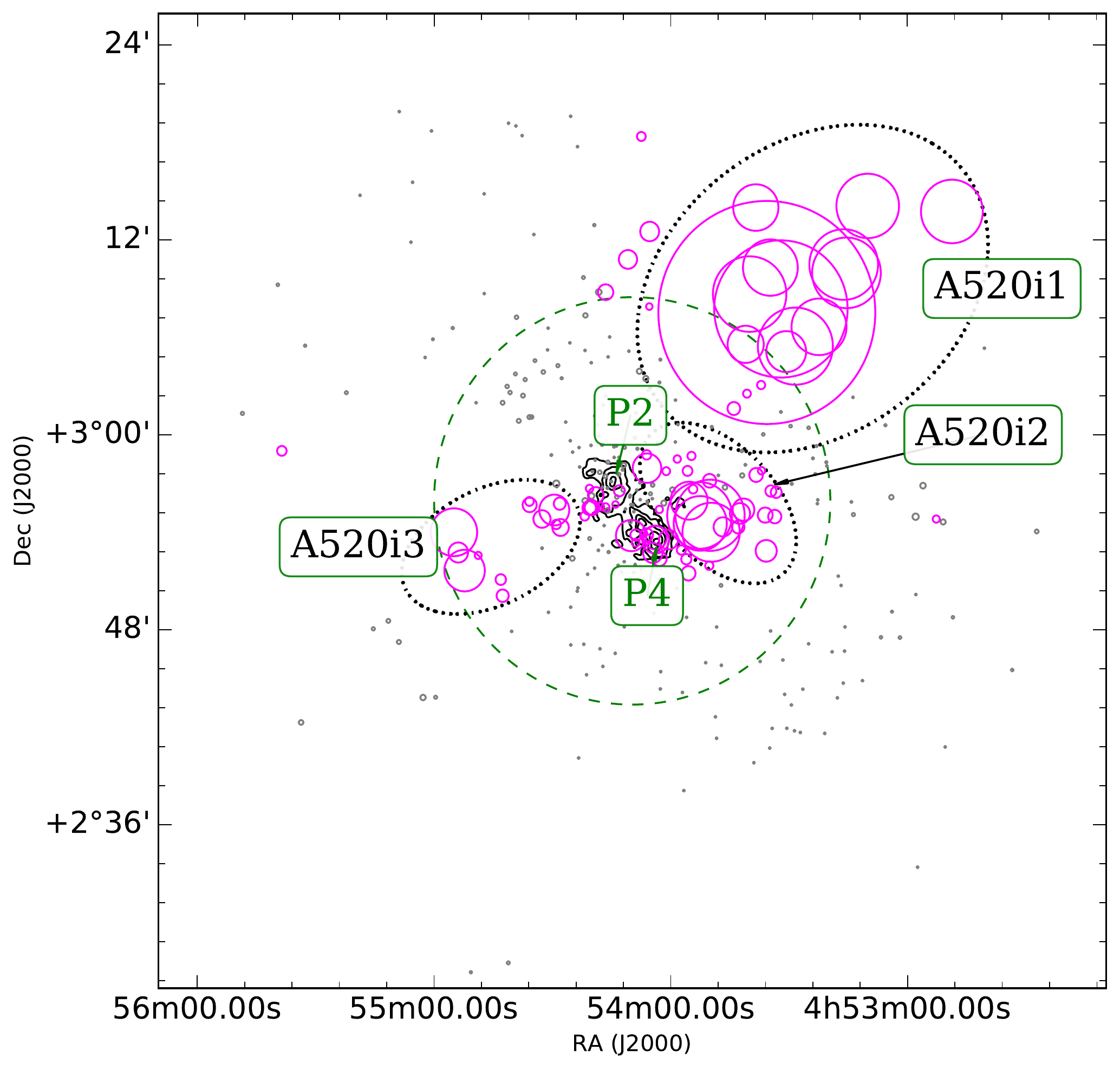}
  \caption{The results from the DS test. Each galaxy is represented with a circle with radius proportional to $e^{\delta}$. The ones with $\delta_i \leq 2.25$ are shown with grey circles, the ones with $\delta_i>2.25$ with magenta circles. The weak lensing mass density contours from \citet{Jee2014} are shown in black and the centres of the main mass concentrations (P2 and P4) are annotated. The green dashed circle has a radius equal to R$_\mathrm{200}$. The recovered structures are outlined with black dotted lines.}
  \label{ds_res}
\end{figure}
                In this Section we use the full data set.

        The unsettled nature of A520 and the presence of substructure is proven beyond doubt by G08 and the numerous weak lensing analyses and X-ray imaging. We employ here the procedure pioneered by \citet{Biviano2002} and adopted by G08 for separating the cluster members according to the probability of them being part of a substructure in order to assess whether their star formation properties depend on this. The first step of this procedure is the Dressler-Shectman test \citep[DS-test,][]{DresslerShectman88}. The test takes every galaxy and its N nearest projected neighbours, calculates their central velocity and velocity dispersion and compares that to the cluster's central velocity and total velocity dispersion to get a deviation ($\delta$), according to the formula:
\begin{equation}
\delta^2 = [(N+1)/\sigma^2] [(\bar{v}_\mathrm{local} - \bar{v})^2 + (\sigma_\mathrm{local} - \sigma)^2].
\end{equation}
        Where $\bar{v}$ and $\sigma$ are the central velocity and velocity dispersion of the cluster as a whole, and the subscript $local$ indicates the values for the group of N+1 galaxies. \citet{DresslerShectman88} used N=10, although $N=n^{1/2}$, with $n$ being the total number of data points, has also been suggested \citep{Bird1994} in order to improve the sensitivity of the DS test to significant substructure, while reducing its tendency to follow fluctuations within the Poisson noise. We followed the latter approach and used $N=n^{1/2}$, which in our case is $N=18$. We run the test on the full data set adopting the previously calculated central velocity and dispersion. 
        The DS test assumes a smooth underlying galaxy distribution upon which some amount of substructure is superimposed. In the case of A520 the presence of a smooth background distribution does not seem obvious. However, because the merger happens very close to the plane of the sky, and the two main merging clumps are indistinguishable in velocity, this is indeed present. This is confirmed by the fact that the global cluster parameters, calculated assuming a
SIS profile, agree with the ones determined by weak lensing. In this case, the galaxies with high probability of being part of a substructure can be considered newcomers to this environment.
 
        The threshold $\delta = 2.25$, above which the galaxies are most likely part of a substructure is determined by running the DS test on 1000 mock cluster catalogues drawn from the parent catalogue of cluster members but with randomly assigned velocities drawn from a Gaussian with the same $\bar{v}$ and $\sigma$ as A520. The distribution of the deviations among all the members of A520 and the 1000 mock clusters is shown in Fig.~\ref{dev_distr}. The point at which the heavy tail in the deviation distribution among A520 members becomes higher than the one of the mock clusters is 
$\delta = 2.25$ . The mock clusters also show a tail towards higher $\delta$ because we only randomise their velocities and not the positions which are also asymmetrical.
        
        Fig.\ref{ds_res} shows the results of the DS test. Each galaxy, from the full data set, is represented with a circle proportional to $e^{\delta}$, where $\delta$ is its probability of being part of a substructure. The galaxies with $\delta>2.25$ are shown in magenta, while the rest are shown in grey. A close group of large magenta circles on this plot indicates the presence of a substructure. The positions of the main mass concentrations from \cite{Jee2014} are annotated. Both P2 and P4 are recovered by the DS test, although the group of magenta circles closest to P2 are somewhat offset from the peak of the black contours showing the mass distribution. The recovery of the central substructures is very sensitive to the choice of central cluster velocity and dispersion. A series of other substructures are also visible, which we have approximately outlined with dotted black lines and named A520i1, i2 and i3. A520i3 likely contains two separate substructures. The one closer to the centre of the cluster is recovered by G08 (the `E' clump on their Fig.11) and by \citet{Jee2014} (clump number 5). The more distant part of A520i3 is outside the area covered by the analysis presented in these two articles. A520i2 is designated `W' by G08. The majority of the galaxies belonging to A520i1 are at $r \gtrsim$R$_\mathrm{200}$ and are not covered by G08 nor by any of the weak lensing analyses. All three substructures outlined in Fig.\ref{ds_res} are aligned with the secondary filament proposed by G08.

\begin{figure}
\centering
\includegraphics[width=\hsize]{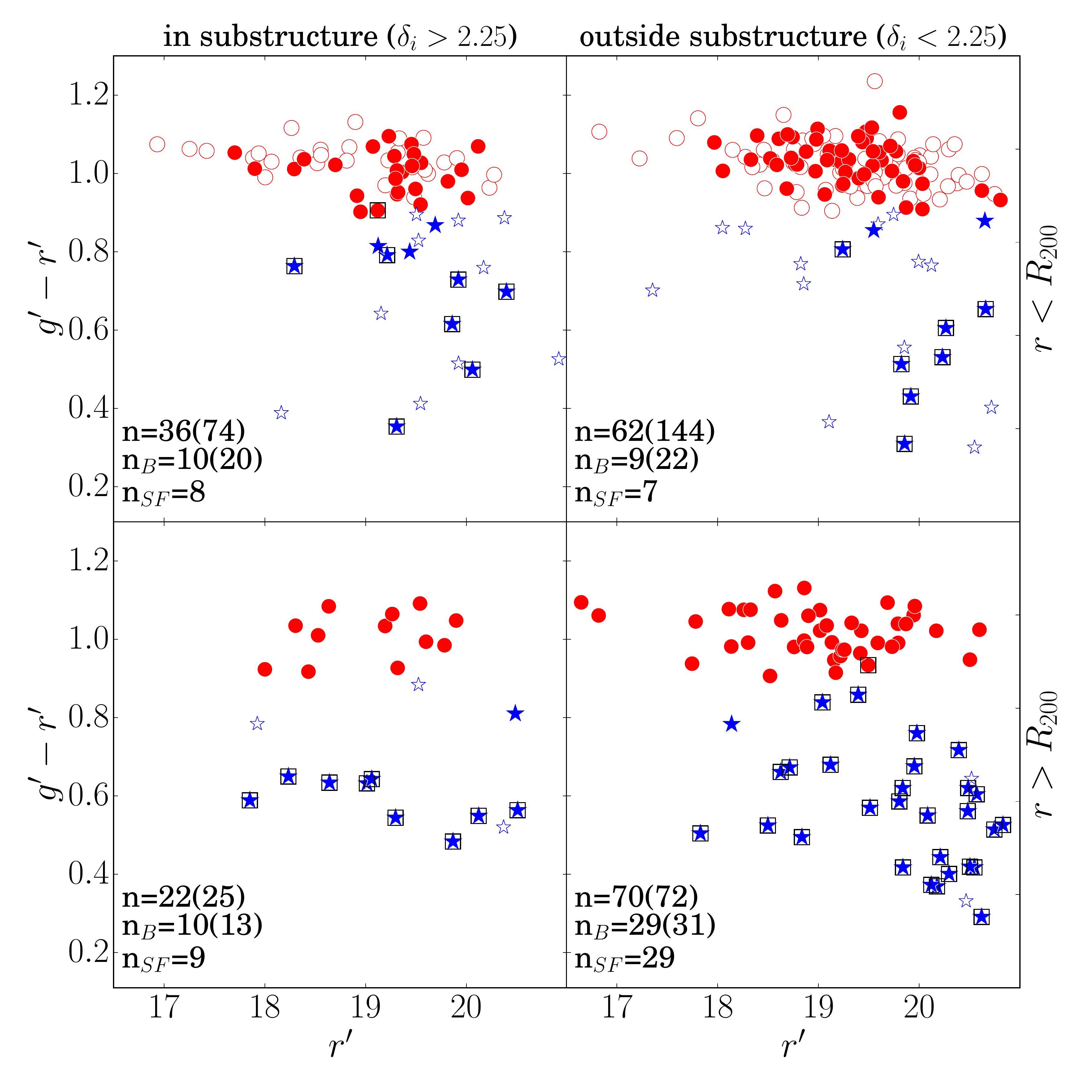}
  \caption{Colour magnitude diagrams for all the galaxies in the full data set. The left column shows galaxies in substructure, the right one outside substructure. The top row shows galaxies within R$_\mathrm{200}$ and the bottom row shows glaxies outside that radius. The galaxies from G08 are shown with empty symbols and the MMT galaxies with full ones. The blue and red galaxies are indicated with blue stars and red circles, respectively. The galaxies with emission lines have a black square around their symbols. The total number of galaxies in each quadrant $n$ is indicated for the MMT data set and in brackets for the full data set. Also indicated are the total number of blue galaxies n$_{B}$ for the MMT data set (full data set), and the total number of star forming galaxies n$_{SF}$.}
  \label{dev_col_mag}
\end{figure}
\begin{figure}
\centering
\includegraphics[width=\hsize]{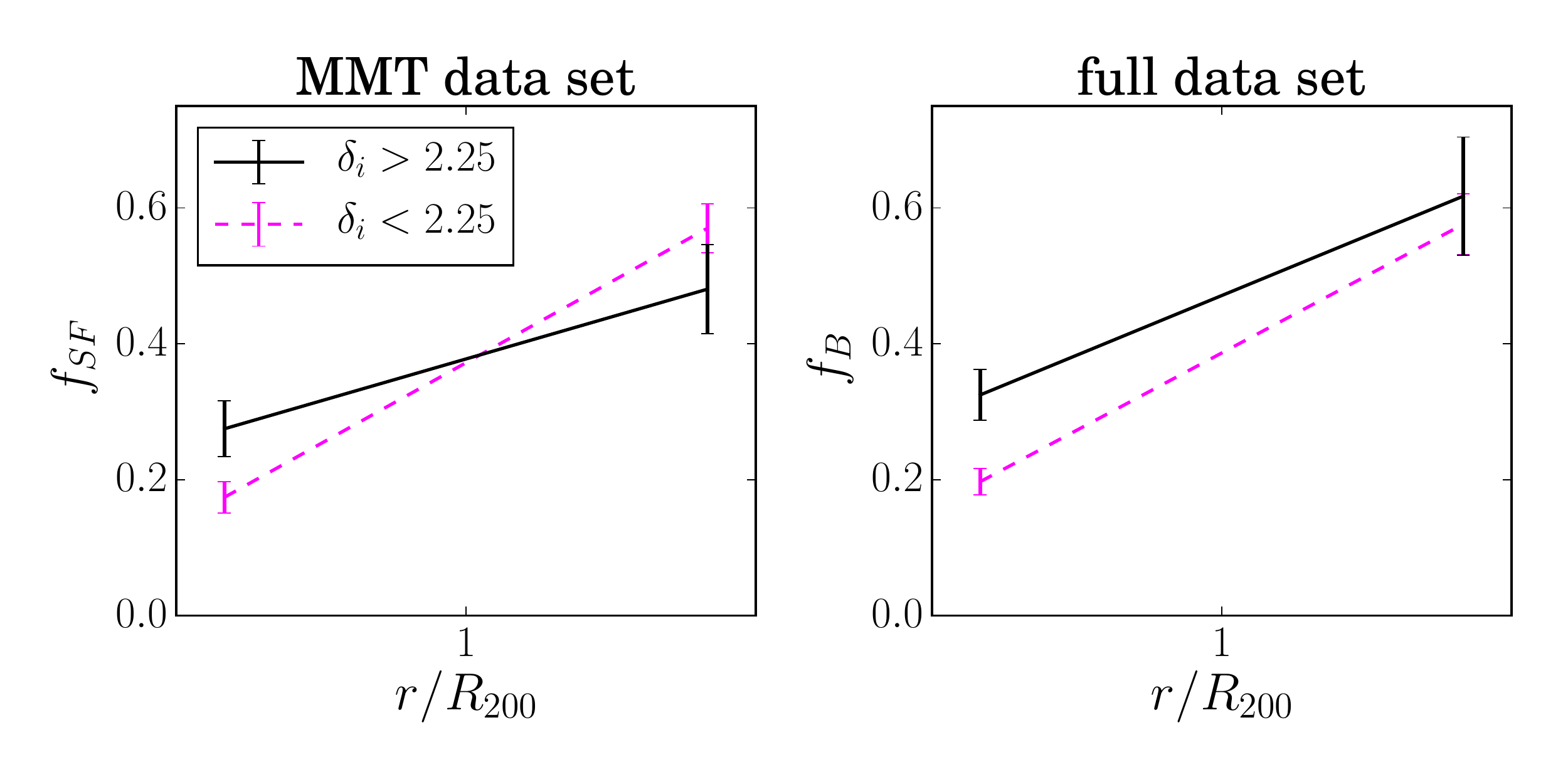}
  \caption{The fraction of star forming (left), and blue (right) galaxies with high (solid, black line), and low (dashed, magenta line) probability of being part of a substructure. The galaxies are separated according to their cluster-centric distance in two bins, r<R$_\mathrm{200}$ and r>R$_\mathrm{200}$. The star forming fraction was determined from the MMT data set, while the blue fraction was determined from the full data set. The points are weighted for completeness.}
  \label{frac_rad_substr}
\end{figure}

        On Fig.~\ref{dev_col_mag} we show the colour-magnitude diagrams of all the members of A520. The filled symbols show the galaxies from the MMT data set. The empty symbols show the G08 galaxies that add up to the full data set. All the star forming galaxies have a black square around their symbol (only available for the MMT data set). The galaxies are split in four quadrants according to their cluster-centric distance and their probability of being part of a substructure. In each quadrant the total number of galaxies in the MMT data set (full data set) is indicated, as well as the number of blue (n$_{B}$) and star forming (n$_{SF}$) galaxies.
        
        On Fig.~\ref{frac_rad_substr}, we compare the fraction of star forming ($f_\mathrm{SF}$), and the fraction of blue ($f_B$) galaxies in the two bins of cluster-centric distance, and for galaxies with high ($\delta_{i}>2.25$) and low ($\delta_{i}<2.25$) probability of belonging to a substructure. The points are weighted for completeness according to their magnitude as shown on Fig.~\ref{2d_compl}c.
        
        As expected, the distance to the cluster centre has by far the strongest effect on both the star forming and blue fractions, which are significantly higher outside R$_\mathrm{200}$ than inside it. The effect of the cluster-centric distance on $f_\mathrm{SF}$ is stronger for galaxies not belonging to a substructure. The much shallower slope of the solid black line on the left panel of Fig.~\ref{frac_rad_substr} shows that the galaxies belonging to substructures already have reduced $f_\mathrm{SF}$ outside R$_\mathrm{200}$. This is in line with the results of the large body of work on group pre-processing \citep[e.g.][]{Balogh1999,McGee2009,Jaffe2016}. Inside R$_\mathrm{200}$ however, this trend is reversed, with the galaxies with $\delta_{i}>2.25$ showing higher $f_\mathrm{SF}$ than the ones with $\delta_{i}<2.25$. This is expected since the substructures indicate recent infall \citep[see also][]{VijayaraghavanRicker2013}. This result also implies that galaxies are not instantly quenched once they pass R$_\mathrm{200}$, as shown by \citet{Haines2015}. The results for the blue fraction $f_B$, shown in the right panel of Fig.~\ref{frac_rad_substr} paint a similar picture inside R$_\mathrm{200}$. Outside R$_\mathrm{200}$ the substructure does not seem to play a role in determining the colours of the galaxies as both galaxies in and out of substructures show statistically identical $f_B$.
        
        A possible explanation for the different trends of $f_\mathrm{SF}$ and $f_B$ with $r$ and $\delta_i$ could be that the time scales on which the star formation is quenched are shorter than the life time of the massive O- and B-type stars that give the blue colours of the galaxies. This quick quenching of star formation however is usually associated with large clusters and ram-pressure stripping. The pre-processing, working in small groups prior to their accretion onto large clusters, is expected to quench star formation on much longer time scales (>2Gyr, \cite{McGee2009}). Although intriguing, we leave the more thorough analysis of the role of pre-processing for a future work; in this article we want to look in detail at the effects that cluster mergers have on their constituent galaxies. We have validated that these results do not vary significantly with a different cut-off value for $\delta_i$.
        
\subsection{The quenching of galaxies in A520}\label{sf_gals}
\begin{figure}
\centering
\includegraphics[width=\hsize]{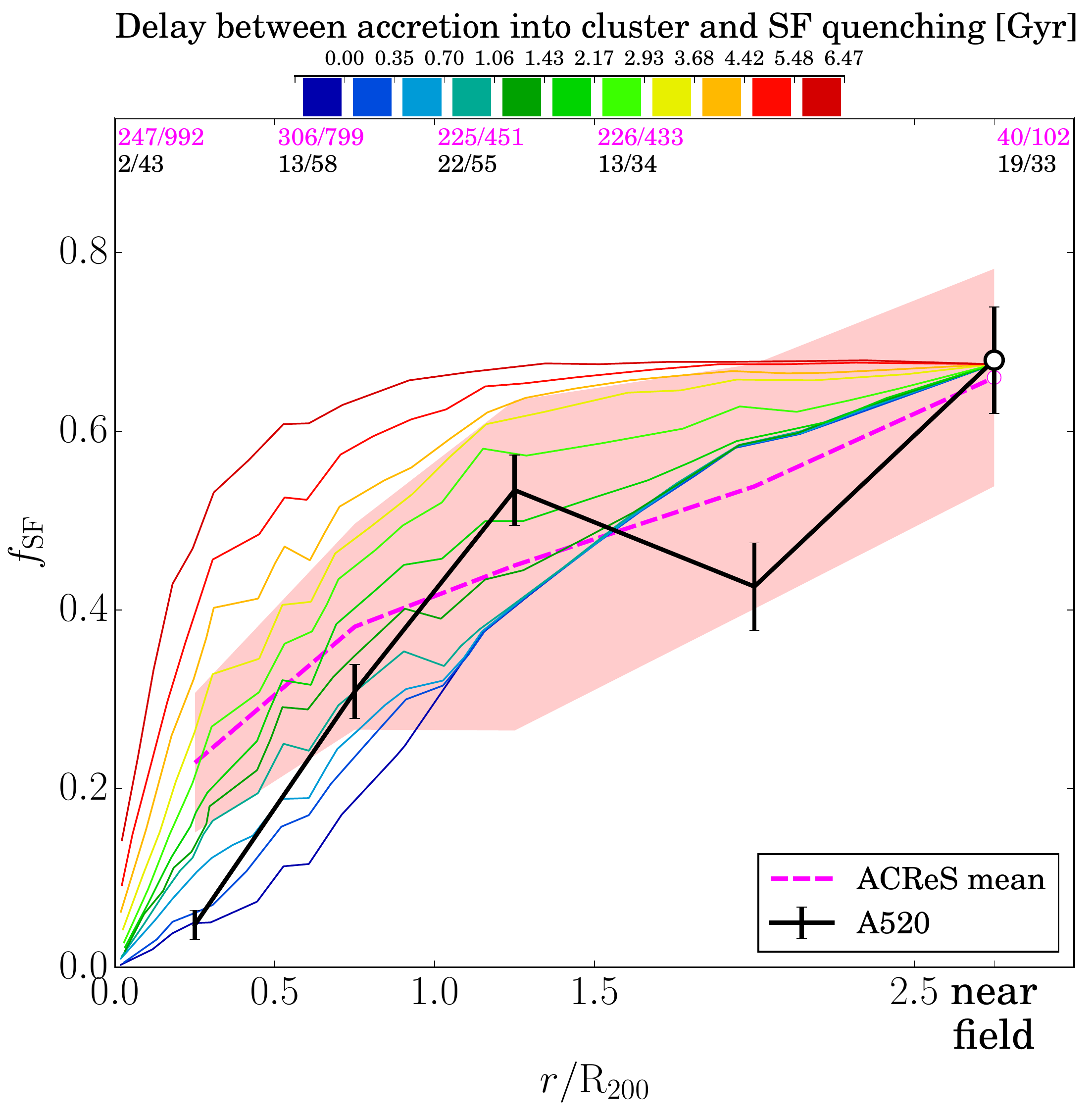}
  \caption{Completeness-corrected fractions of star forming galaxies as a function of cluster-centric distance. The black solid line shows A520. The thick dashed magenta line shows the mean of the ten clusters from the comparison sample. The region shaded in magenta shows the spread ($\pm$1$\sigma$) among the individual clusters. The outermost point, indicated with a large circle, shows $f_\mathrm{SF}$ in the near field region. The last radial bin is double the size of the inner ones due to a low number of points. The numbers in each radial bin show the number of galaxies in that bin before correction for completeness (number of star forming galaxies $N_\mathrm{SF}$ / total number of galaxies $N_\mathrm{all}$). The magenta and black numbers show the comparison sample and A520, respectively. The thin lines of different colour show the expected $f_\mathrm{SF}$-radius relation from observations of 75 massive clusters in the Millenium simulation. The star formation in the infalling galaxies is quenched upon passing through R$_\mathrm{200}$ of the cluster for the first time, with a delay indicated by the different colour, according to the colour bar on top of the plot. $f_\mathrm{SF}$ in the infall region of the simulated clusters is normalised to the median $f_\mathrm{SF}$ in the near field of the ten comparison clusters.}
  \label{em_line_frac}
\end{figure}
        In this section and in Section~\ref{gal_distr}, we use only the MMT data set, since we have no information on the presence and strength of the emission lines from G08. For consistency, we only use the ACReS observations of the clusters from the comparison sample. 
        
        Fig.~\ref{em_line_frac} shows the fraction of star forming galaxies $f_\mathrm{SF}$ among the cluster members of A520 and the comparison sample in bins of cluster-centric distance normalised by the characteristic size of each cluster. The thick, dashed, magenta line represents the median $f_\mathrm{SF}$-r relation for the ten clusters in the comparison sample. The region shaded in magenta shows the spread among the individual clusters. The black solid line with error bars shows the results for A520. The errors are calculated assuming binomial distribution. The points are completeness corrected, based on their $r'$ magnitudes, taking into account the completeness shown on Fig.~\ref{2d_compl}\textit{c}. The number of galaxies in each bin, before completeness correction, is indicated at the top of the Figure. Black numbers are used for A520 and magenta for the comparison sample. The number of star forming galaxies and the total number of galaxies are indicated $N_\mathrm{SF}$ / $N_\mathrm{all}$. The last point shows $f_\mathrm{SF}$ in the near field region as defined in Fig.~\ref{SIS}.
        
        From 0.7R$_\mathrm{200}$ (${\sim}$R$_\mathrm{500}$) out to the near field region, A520 is indistinguishable from the comparison sample. Closer to the centre A520 differs significantly, showing a lower $f_\mathrm{SF}$ than all of the comparison clusters. For A520, $f_\mathrm{SF}=$0.05$\pm$0.02 in the innermost bin. The comparison sample has $f_\mathrm{SF}=$0.23$\pm$0.08 in the same bin. The $f_\mathrm{SF}$ for the different clusters in the first bin are between 0.13 (A383) and 0.42 (A963). This result agrees well with the estimated blue fraction by \cite{Butcher&Oemler1984}, although the methods used are completely different.
        
         The thin solid lines of different colour represent the expected $f_\mathrm{SF}$-radius relation if the infalling galaxies are quenched with a certain delay after their first crossing of R$_\mathrm{200}$. These lines are adapted from \citet{Haines2015} and are based on 75 massive clusters from the Millennium simulations \citep{Springel2005}. The $f_\mathrm{SF}$ of the infalling galaxies in the simulated clusters is normalised to the median $f_\mathrm{SF}$ of the galaxies in the near field of the ten comparison clusters. As concluded by \citet{Haines2015} the data of all the clusters are consistent with the star formation continuing in the infalling galaxies for $\sim$2~Gyr after accretion. In contrast, the central parts of A520 are more consistent with the presence of an additional quenching mechanism. This region of the cluster is populated by galaxies that have experienced the high-velocity core passage, and are supposedly most affected by the merger. This is also the region occupied by the hot X-ray emitting ICM.
        
        We have confirmed that this result does not change significantly if the weights of each data point are based on the colour dependent completeness, or removed altogether. In the latter case the difference between A520 and the comparison sample is greater and extends out to 2R$_\mathrm{200}$.
        
        It's worth mentioning that our survey, as any magnitude limited survey, under-represents low mass, passive galaxies due to their high M/L ratio (see Fig.\ref{d4_stmass}). If we are missing any galaxies they are most likely passive, as at any given stellar mass, a star forming galaxy would be brighter. We note that the comparison sample doesn't suffer from such a bias, as its infra-red-based target selection makes it, in effect, stellar-mass limited rather than magnitude limited. Sampling A520 and the ACReS clusters on equal footing would likely show even greater difference between their respective $f_\mathrm{SF}$ - radius relations. The assumption that a merging cluster of galaxies is well represented by an Isothermal Sphere (\ref{glob_params}) cannot possibly affect the results concerning the centre of the cluster.

\subsection{Distribution of passive, star forming, and recently quenched galaxies}\label{gal_distr}        
\begin{figure*}
\centering
\includegraphics[width=\hsize]{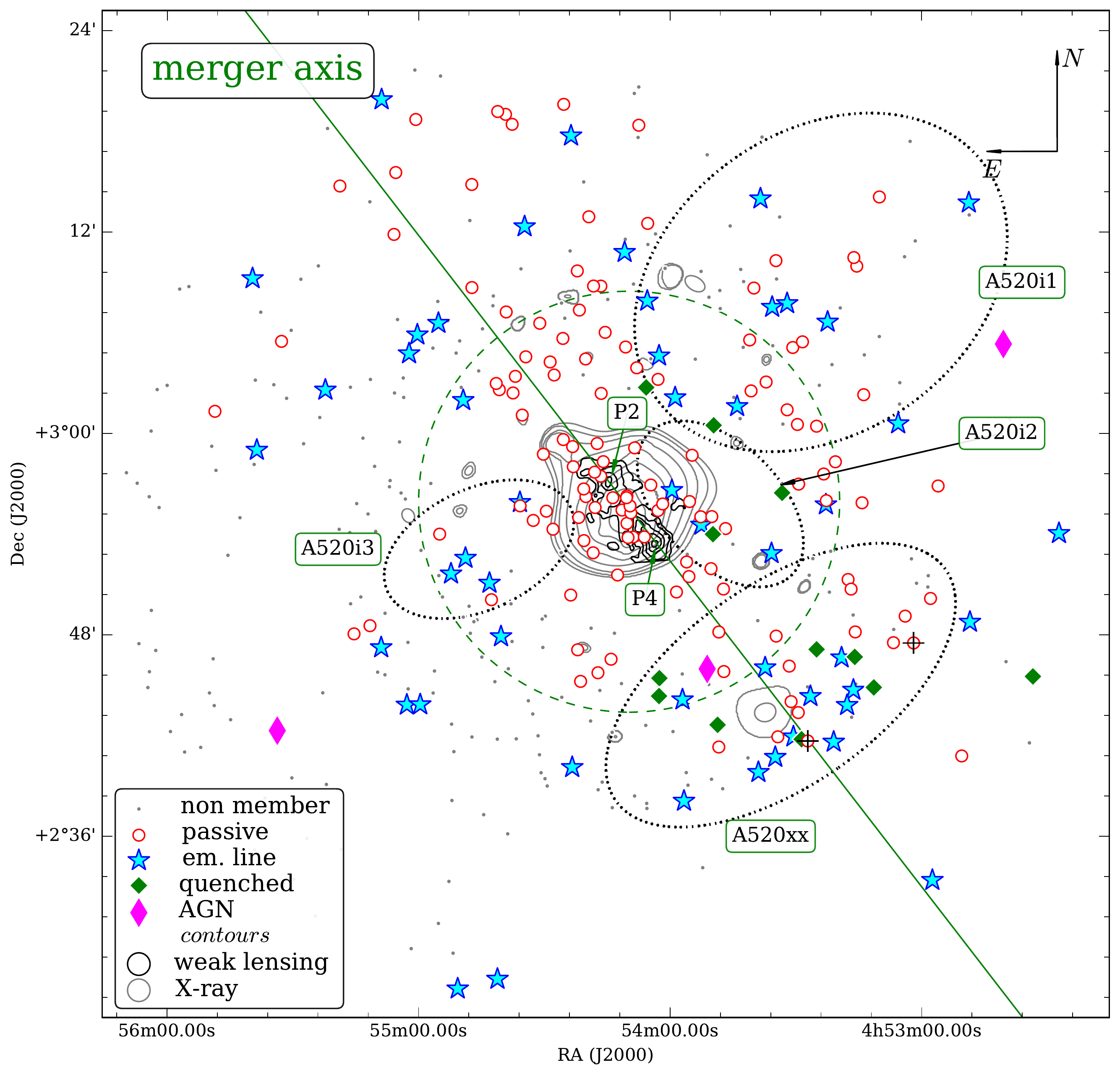}
  \caption{Distribution on the sky of all galaxies from the MMT data set. The small grey dots are galaxies with $z$; not members of A520. All the other symbols show A520 members. The red circles are passive galaxies. The large magenta diamonds are AGNs and the cyan stars are star forming  galaxies. The small green diamonds are recently quenched galaxies. The black contours in the centre show the mass distribution from \citet{Jee2014}, the grey contours show the X-ray emission from XMM Newton. The green dashed circle has a radius equal to the R$_\mathrm{200}$ of the cluster. The solid green line connects the P2 and P4 mass concentrations from \citet{Jee2014} and indicates the approximate orientation of the merger axis. The black dotted ellipses with name tags are the DS recovered substructures, the same as in Fig.\ref{ds_res}. The addition of A520xx is discussed in the text.}
  \label{SFR_on_the_sky}
\end{figure*}

        Fig.~\ref{SFR_on_the_sky} shows the distribution of our targets on the sky with respect to the weak-lensing-derived mass distribution (black contours, \cite{Jee2014}) and X-ray emission (grey contours). The small grey points are galaxies from the MMT survey that are not members of A520. 
Other symbols show the galaxy classifications as in Fig.~\ref{d4_stmass}, with the addition of the thin magenta diamonds showing the three cluster members with AGN contribution to their spectra. The green dashed circle shows the R$_\mathrm{200}$ of the cluster. The green solid line goes through the P2 and P4 mass concentrations from the weak lensing map, thus indicating the axis along which the merger progresses. This should be taken as the approximate axis of the merger as we cannot rule out some impact parameter projected on the plane of the sky. Presumably this is also the axis of the filament along which P2 and P4 were moving, and the potential direction for new accretion. Overplotted with black dotted lines are the three regions containing substructure, identified in Fig.~\ref{ds_res}, plus a fourth region called A520xx, which we discuss below.

        The distribution of the galaxies on this plot is asymmetrical. All cluster members lying to the South-East of the core of the merger (left-down) form a relatively narrow structure elongated towards the centre of A520, and are predominantly star forming. A520i3 is part of the extension of this structure inside R$_\mathrm{200}$. On either side of this structure there are a number of galaxies covered by our MMT survey which are not part of the cluster. This structure, together with the groups A520i2, within R$_\mathrm{200}$, and A520i1 at higher cluster-centric radius, further strengthen the impression of a secondary filament, as proposed by G08.

        Within R$_\mathrm{200}$ there is also strong asymmetry in the distribution of passive and star forming galaxies. Along the merger axis there is a stripe of passive galaxies that extends outside R$_\mathrm{200}$ to the North-East, along the axis of the main filament. To the South-West however this stripe of passive galaxies is truncated roughly at R$_\mathrm{200}$ by a concentration of galaxies, dominated by star forming and recently quenched galaxies. We refer to this concentration as A520xx and have outlined it, by eye, with a dotted, black line on Fig.\ref{SFR_on_the_sky}. The velocity distribution of the galaxies within this group is not dissimilar to that of the cluster itself, which could be the reason why it was not detected by the DS test. This group contains ${\sim}~67\%$ of the recently quenched galaxies found in A520, and approximately half of the galaxies within this group are either star forming or recently quenched. Among them are also the two galaxies with the highest stellar mass in our survey (we did not target the two cD galaxies of A520), both showing unusually low $D_n(4000)$ for their stellar mass (see Fig.~\ref{d4_stmass}), although higher than the average value of the star-forming comparison sample. These are indicated with black crosses on Fig.~\ref{SFR_on_the_sky}. 
        
        Coinciding with A520xx is a tentative detection of extended X-ray emission. This X-ray detection likely does not contain all the emission as it lies at the edge of XMM's field of view, although it is significantly above the, usually elevated, noise there. It is also likely that our MMT survey does not target all the galaxies from this potential infalling group, as there are no galaxies targeted in this direction farther than $r{\sim}$1.5R$_\mathrm{200}$. The distribution of extended objects in our photometric catalogue shows increased density in this region. Further observations are needed to test if A520xx is indeed a group or a cluster infalling along the same filament as P2, and to pinpoint the quenching mechanism affecting these galaxies.
        
        The other ${\sim}33\%$ of the recently quenched galaxies are well within R$_\mathrm{200}$ but are all grouped to the north-west of the merger centre, along the secondary filament, as part of the substructures A520i1 and A520i2. This secondary filament is also the axis along which the star forming galaxies reach closest to the merger centre.
\begin{figure*}
\centering
\includegraphics[width=1.0\hsize]{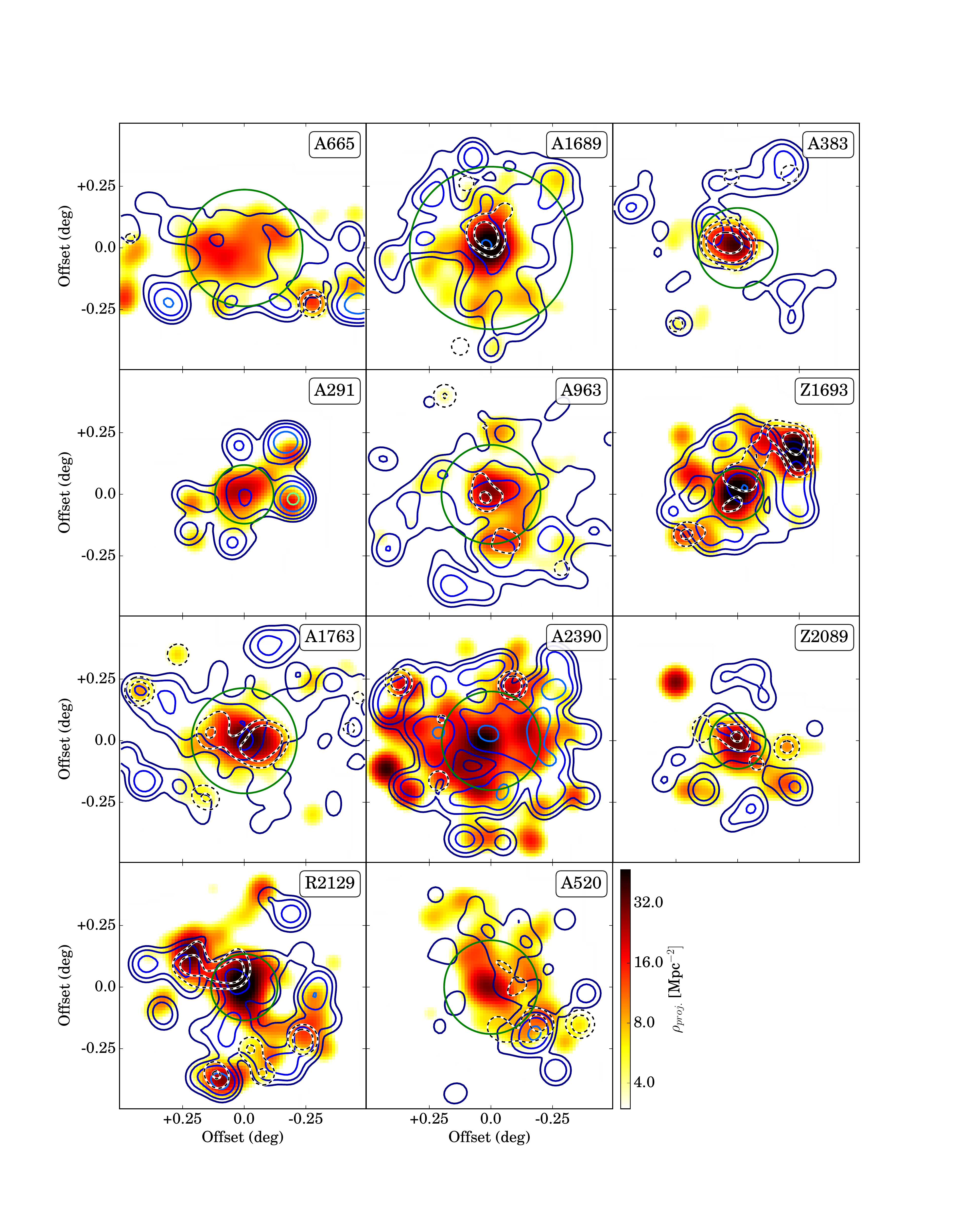}
  \caption{The projected number density of all members of the comparison clusters and A520. These are completeness corrected \textit{kde} densities estimated separately for passive, active and recently quenched galaxies using the same smoothing kernel size. $log(\rho_\mathrm{proj})$ of the passive galaxies is shown as an underlying image, of the star forming galaxies with solid contours, and of the recently quenched ones with dashed contours. The contours are drawn at the levels indicated with the tick marks on the colour bar. The green circle has a radius equal to R$_\mathrm{200}$ of the given cluster.}
  \label{proj_density}
\end{figure*}

        Fig.~\ref{proj_density} shows the projected number density of the passive, star forming, and recently quenched members of the ten clusters from the comparison sample and A520. The underlying image shows the number density of passive galaxies. The number density of the star forming galaxies is shown by the overlaid solid contours. The number density of recently quenched galaxies is shown with the dashed contours. The points on the Figure are weighted for completeness, according to Fig.~\ref{2d_compl} a). In all the comparison clusters the peaks of the densities of passive and star forming galaxies are close to each other and in most cases they coincide. Even though in all the clusters the fraction of star forming galaxies drops towards their respective centres (Fig.~\ref{em_line_frac}), their density still increases toward the central regions. This is partially due to the large overall increase in total galaxy density but also because, as shown by \citet{Haines2015}, galaxies are not instantly quenched upon accretion. Possible exceptions are A2390 and A665, where the main peak of the density of passive galaxies does not have a corresponding peak in the density of star forming galaxies. Neither of these clusters appear relaxed on Fig.\ref{proj_density} confirming earlier suggestions of their unsettled nature by \cite{Abraham1996} and \cite{Markevitch2001, Dasadia2016}, respectively. This supports the idea that star formation is being rapidly quenched in galaxies within merging clusters, out to R$_\mathrm{500}$.
                
        Although the number of recently quenched galaxies per cluster is relatively low ($N_\mathrm{rq}\le38$, $\tilde{N}_\mathrm{rq} =\sim$16) they reveal a similar picture. If there are any such galaxies in the given cluster, their densities also tend to peak around the overall peak in galaxy density. A similar result is also found in the analysis of the distribution of post-starburst galaxies in local clusters by \citet{Paccagnella2017}. This similarity is encouraging, since their selection of post-starburst galaxies is based on the absence of emission lines and the presence of $H_\delta$ in absorption and thus differs from our recently quenched galaxies. The exceptions from this picture are again A2390 and A665 where the peaks in density of recently quenched galaxies are found outside R$_\mathrm{200}$. We found no recently quenched galaxies in A291.
        
        We compare this to the density distribution in A520, shown in the last panel. The passive galaxies distribution is elongated along the merger axis (same orientation as Fig.~\ref{SFR_on_the_sky}). This density peak however is accompanied by a complete absence of star forming galaxies. This is not the case with the density peak associated with A520xx visible in the lower-right part of the plot. From the  clusters presented here, A520 is the only one to show such an extensive depletion of star forming galaxies in its core. We note that many of the comparison clusters show some secondary concentrations of galaxies of all types (e.g. A963, R2129, A1693) yet none show such a severe drop in the density of star forming galaxies at their centre.
\section{Discussion}\label{discussion}

        The merging clusters of galaxies are a very specific kind of environment which was probably much more common at high redshifts \citep{Mann&Ebeling2012}, when the massive early-type galaxies found in the local clusters were formed, and the bulk of the red sequence established. Although most clusters of galaxies show some amount of substructure \citep[][this work]{DresslerShectman1988, JonesForman1999}, the merging clusters of galaxies that show a clear offset between the ICM and the galaxy concentrations are relatively rare at low $z$ \citep{MaEbeling2010}. While the importance of the accretion of smaller structures onto clusters is well appreciated \citep{McGee2009}, the effects of mergers of equal mass clusters are still poorly understood. We present here an analysis of the distribution of star formation activity in one of the lowest redshift merging clusters known. We find the core of the cluster to be depleted of any star formation, with the "passive" region being extended along the merger axis. The fact that this gap in the star forming galaxies distribution is elongated along the merger axis hints at a possible causal connection between the two. This absence of star forming galaxies in the centre also suggests very little accretion of galaxies along an axis aligned with the line of sight, which would appear projected on top of the merger and would then fill in the gap.
        
        When discussing the results from the analysis presented in Section~\ref{results} we have to keep in mind that we are only measuring star formation in the very centres of the galaxies, as discussed in Section~\ref{gal_class}. Environment-related processes affect the star formation in galaxies indirectly, by altering the amount and distribution of cold gas within them. These processes are likely to have larger effects in the outskirts of the galaxies where the galaxy's potential well is shallower \citep{TonnesenBryan&vanGorkom2007}. Because we are only measuring the star formation properties in the central regions of the galaxies, where the star formation might continue long after its outskirts are experiencing the effects of the environment, we only classify galaxies as passive when their star formation is completely quenched and their transformation to passive galaxy is completed. Galaxies that are in the process of being transformed but still have some residual star formation are likely to host it in their respective centres and hence will be classified as star forming by our survey. If the environmental processes are inducing star formation by compressing the available cold gas, this is also likely to happen in the central parts of the galaxies \citep{HendersonBekki2016}. With that in mind, the core of A520 shows no signs of merger-induced star formation, neither ongoing nor recent.
        
        Using the velocity of the merger (2300 kms$^{-1}$, \citet{Markevitch2005})  and the current separation on the sky between P2 and P4 we estimate a time since core passage of $\sim$0.2-0.4~Gyr, assuming an inclination with regards to the plane of the sky of 30$^{\circ}$ or 60$^{\circ}$. If any star formation has taken place in the central galaxies of A520 on these time scales, this would leave a trace in their $D_n(4000)$ values. We do not observe such traces. Using a relation between the age of the stellar populations and $D_n(4000)$ \citep{Lopez2016}, based on data from the CALIFA survey, we estimate the ages of the recently quenched galaxies to be all >1~Gyr, with a median age of $\sim$3~Gyr. The age of the passive galaxy population determined in the same way is $\sim$6~Gyr. Analysis of the stellar population ages based on the summed spectra presented in Fig.~\ref{all_sp} shows somewhat younger ages than the ones calculated from $D_n(4000)$, although with a similar offset between the two galaxy types. The {\sc Ca\,i\,i} index \citep{LeonardiRose1996}, defined as {\sc Ca\,i\,i} H+H$\epsilon$/{\sc Ca\,i\,i} K, for the passive galaxies spectra shown in Fig.~\ref{all_sp} is 1.26. The same index for the recently quenched galaxies is 0.86. The analysis presented by \citep{LeonardiRose1996} shows that these values are consistent with the galaxies not having any star formation in the last $\gg$2~Gyr, in the case of the passive galaxies, and $\lesssim$1~Gyr in the case of the recently quenched ones. The estimated lifetime of the recently quenched phase derived in Section~\ref{rec_q} is lower than both of these ages, although all of them are greater than the time since core passage in A520. There could be a number of reasons for those different ages; for example we used a simplified model as opposed to the observationally calibrated relations used above, and metallicity also has an effect on D4000.
        
        Many of the studies dedicated to galaxy properties in merging clusters of galaxies observe an increase in both star formation and AGN activity, starting with the classical work by \cite{Caldwell1993}, and \cite{CaldwellRose1997} and continuing with more recent studies by \cite{Ferrari2005, HwangLee2009, ChungSM2009, Sobral2015, Stroe2014, Stroe2016}. In contrast, by observing the infrared properties of the galaxies in A2255, \citet{Shim2011} conclude that the cluster mergers are suppressing star formation in member galaxies. \citet{Pranger2014} also observe a concentration of recently quenched galaxies in the vicinity of the core of a merger, but not any ongoing star formation. \cite{Marcillac2007} also concludes that the star forming galaxies in RX J0152.7-1357 are newly infalling ones, rather than being affected by the ongoing cluster merger. The distribution of star forming and recently quenched galaxies in A520 does not suggest the presence of a recent star burst. Instead, the data suggest a quenching of the star formation upon infall at around R$_\mathrm{200}$. We cannot rule out a short burst of star formation preceding its quenching. If the members of P2 and P4 have experienced star formation quenching as far out as R$_\mathrm{200}$ this could explain the lack of signatures of any recent star formation as, from geometrical considerations, this would have happened more than 1~Gyr prior to the core passage. By now the old stellar population will completely dominate the light from those galaxies and the luminosity-weighted age will be the age of these underlying old stars.

        Assuming that the two merging components of A520 are well represented by our comparison sample (dashed magenta line on Fig.~\ref{em_line_frac}), then the sharp decrease in the fraction of star forming galaxies towards the centre of A520 is indicating a rapid quenching of the star formation in the galaxies that have gone through the hot intra-cluster gas at the high velocities typically associated with cluster mergers. The fact that most of the star forming and recently quenched galaxies within R$_\mathrm{200}$ are grouped in a separate substructure indicates that they are newly infalling. The galaxies that have experienced the core passage are more consistent with instant quenching (thin blue line on Fig.~\ref{em_line_frac}) and don't show any signs of recent or ongoing star formation \citep{Haines2015}. 
        
        There are few physical mechanisms that act on galaxies infalling towards relaxed clusters that could be enhanced in merging clusters. Tidal interactions with the clumpy gravitational potential of the merging clusters could potentially displace the loosely bound outskirts of the galaxies and make the gas there more susceptible to ram pressure stripping. We are observing a depletion of star formation in the very centres of the galaxies in the core of A520. This could indicate the presence of a different mechanism in addition to the tidal force. 
        
        \citet{Stroe2015} find an increased density of star forming galaxies in the "Sausage" merging cluster. They point at the shock waves travelling through the intra-cluster medium as the reason for this. A separate analysis by \citet{Stroe2015a} actually detects 21cm emission from neutral hydrogen, coming from members of the merging cluster. While this is at odds with a number of studies showing that cluster galaxies are usually devoid of {\sc H\,i} \citep[e.g.][]{Solanes2001,Verheijen2007,Deshev2009,Chung2009}, it is a requirement for ongoing star formation. Shocks have been observed in A520 \citep{Markevitch2005,Vacca2014,Wang2016} but we do not find any star formation associated with them. 
        
        The analysis presented here cannot pinpoint the relative contribution of different physical mechanisms to the evolution of the galaxies in A520. However, because the central parts of A520 show different $f_\mathrm{SF}$ from the comparison sample (Fig.~\ref{em_line_frac}) we can conclude that the cold gas content of the infalling galaxies was either stripped at high cluster-centric radii, or was stripped quickly and efficiently without any star formation taking place closer to the cluster centre. A further effort into simulating ram pressure stripping in the presence of tidal interactions could help us understand the history of the galaxies in A520. Processes acting prior to accretion onto the main cluster, such as pre-processing, cannot possibly introduce the different $f_\mathrm{SF}$-radius relations shown on Fig.~\ref{em_line_frac}, as they would alter the curves over the entire observed range of cluster-centric distances.

        The recently quenched galaxies belonging to A520i1 and A520i2 are all within R$_\mathrm{200}$ of the cluster, and their presence there could be due to the interaction with the hot intracluster matter. The majority of those belonging to A520xx however are outside R$_{200}$, where the ICM density is low, and the shock front associated with the cluster merger is yet to reach. If the transformation of these galaxies is associated with interaction with A520, this could indicate the presence of a significant amount of warm-hot intergalactic medium, which, combined with high relative velocities, can exert strong ram pressure. An alternative explanation of pre-processing within smaller groups or filament needs to account for the quenching taking place at the doorstep of a big cluster, as the lifetime of this phase is $\lesssim$0.5~Gyr. Perhaps the increased local density which the infalling galaxies would encounter for the first time can explain this concentration of recent or ongoing star formation \citep{Mahajan2012}. If future observations confirm A520xx as an infalling rich group, this could potentially imply that the interaction with A520 at r>R$_\mathrm{200}$ could also trigger star formation, as the $\sim$50\% fraction of star forming galaxies is higher than the one usually found in groups or clusters.

\section{Summary}\label{summary}
This article contains the results of a case study of the effects that merging clusters of galaxies have on their constituent galaxies. We present new spectroscopic data of galaxies in the merging cluster A520 at $z=0.2$. The new 363 unique measured redshifts are added to the 293 published by G08 to estimate global cluster parameters and cluster membership. We find 315 members in total, 148 of which are new. We split this sample into `star forming' and `passive' based on the presence / absence of emission lines in their spectra, and examine their spatial distribution when compared to that in relaxed clusters at the same redshift. We also identify a population of recently quenched galaxies constituting objects with no emission lines but with $D_n(4000)$ values lower than the mean value of the star forming galaxies with the same stellar mass. The main results can be summarised as:
\begin{itemize}
  \item The central ${\sim}1.5$Mpc in A520 (${\sim}$R$_\mathrm{500}$) is almost completely depleted of star forming galaxies. This is in contrast to the star forming fraction in a comparison sample of mostly regular galaxy clusters. This difference is found only in the central region of the cluster, with the outskirts and the near field region of A520 showing no deviation from the comparison sample.
  
  \item Projected on the sky, the region in A520 depleted in star forming galaxies is elongated along the merger axis, hinting at a causal connection between the two. 
  
  \item There is no cluster among our comparison sample that shows such a gap in the density of star forming galaxies like the one observed in A520. All of them show centrally peaked density distribution of both passive and star forming galaxies.
  
  \item Converting D4000 into luminosity weighted stellar population age, we find no signs of star formation within the centres of the galaxies populating the two merging cores of A520, on the time scales of the merger.  
  \item We report the tentative discovery of an infalling group from the south-west along the merger axis.
  
  \item This infalling group is rich in star forming galaxies and also contains ${\sim}67\%$ of the recently quenched galaxies in our data set.
  
  \item The earlier-proposed second filament feeding matter into the merger is also apparent in our data.
  
  \item This secondary filament possibly contains most of the star forming galaxies within R$_\mathrm{200}$ of A520 and the remaining ${\sim}33\%$ of the recently quenched galaxies.
\end{itemize}
\begin{acknowledgements}
We thank the referee, Yara Jaff\'e, for the numerous ideas and suggestions which helped improve this article. We thank Michael Balogh and Jaan Pelt for the fruitful discussions during the data analysis. 
\\
        This work was supported by institutional research funding IUT26-2 and IUT40-2  of the Estonian Ministry of Education and Research, by the Centre of Excellence “Dark side of the Universe” (TK133) financed by the European Union through the European Regional Development Fund. Part of this work was also financed by DoRa programme activity no. 6, supported by European Social Fund and Archimedes foundation, and by "Jaan Einasto Scholarship".
        
        The work presented here was also supported by K-GMT Science Program (PID: 14B-MMT001/2014B-UAO-G3) funded through Korean GMT Project operated by Korea Astronomy and Space Science Institute (KASI).
\\
This research used the facilities of the Canadian Astronomy Data Centre operated by the National Research Council of Canada with the support of the Canadian Space Agency.
        This research made use of the ``K-corrections calculator'' service available at http://kcor.sai.msu.ru/
\end{acknowledgements}
\bibliographystyle{aa} 
\bibliography{/home/tazio/works/references} 
\end{document}